\documentclass[journal]{IEEEtran}

\usepackage{setspace}
\usepackage{lipsum}

\usepackage{graphics,
           psfrag,
           epsfig,
           amsthm,
           cite,
           amssymb,
           url,
           dsfont,
           subfigure,
           algorithm,
           algorithmic,
           balance,
           enumerate,
           color,
           setspace
}
\usepackage{amsmath}

\usepackage{epstopdf}

\newcommand{\eref}[1]{(\ref{#1})}
\newcommand{\sref}[1]{Section~\ref{#1}}

\newcommand{\fref}[1]{Figure~\ref{#1}}

\newcommand{\cref}[1]{Constraint~\ref{#1}}

\newcommand{\tref}[1]{Table~\ref{#1}}
\newcommand{\algref}[1]{Algorithm~\ref{#1}}

\newcommand{\ignore}[1]{}

\IEEEoverridecommandlockouts
\begin{document}

\title{\vspace{-.5cm}{A Tutorial on Clique Problems in \\Communications and Signal Processing}}

\author{
   \IEEEauthorblockN{Ahmed Douik, \textit{Student Member, IEEE}, Hayssam Dahrouj, \textit{Senior Member, IEEE},\\ Tareq Y. Al-Naffouri, \textit{Senior Member, IEEE}, and Mohamed-Slim Alouini, \textit{Fellow, IEEE}\vspace{-.8cm}}

\thanks {Ahmed Douik is with the Department of Electrical Engineering, California Institute of Technology, Pasadena, CA 91125 USA (e-mail: ahmed.douik@caltech.edu).

Hayssam Dahrouj is with the Department of Electrical Engineering, Effat University, Jeddah 22332, Saudi Arabia (e-mail: hayssam.dahrouj@gmail.com).

T. Y. Al-Naffouri and M.-S. Alouini are with the Division of Computer, Electrical and Mathematical Sciences, and Engineering, King Abdullah University of Science and Technology, Thuwal 23955-6900, Saudi Arabia (e-mail: \{tareq.alnaffouri,slim.alouini\}@kaust.edu.sa).
}
}

\maketitle

\begin{abstract}
\textcolor{black}{Since its first use by Euler on the problem of the seven bridges of K{\"o}nigsberg, graph theory has shown excellent abilities in solving and unveiling the properties of multiple discrete optimization problems. The study of the structure of some integer programs reveals equivalence with graph theory problems making a large body of the literature readily available for solving and characterizing the complexity of these problems. This tutorial presents a framework for utilizing a particular graph theory problem, known as the clique problem, for solving communications and signal processing problems. In particular, the paper aims to illustrate the structural properties of integer programs that can be formulated as clique problems through multiple examples in communications and signal processing. To that end, the first part of the tutorial provides various optimal and heuristic solutions for the maximum clique, maximum weight clique, and $k$-clique problems. The tutorial, further, illustrates the use of the clique formulation through numerous contemporary examples in communications and signal processing, mainly in maximum access for non-orthogonal multiple access networks, throughput maximization using index and instantly decodable network coding, collision-free radio frequency identification networks, and resource allocation in cloud-radio access networks. Finally, the tutorial sheds light on the recent advances of such applications, and provides technical insights on ways of dealing with mixed discrete-continuous optimization problems.}
\end{abstract}


\section{Introduction}

\textcolor{black}{The recorded history of graph theory dates back to the 1700s when the Swiss mathematician Leonhard Euler learned about the intriguing problem of the seven bridges of K{\"o}nigsberg. The solution proposed by Euler required an innovative level of abstraction, later named by James Joseph Sylvester in his 1878 \emph{Nature} paper \cite{1878NaturS} as ``graphs". While most of K{\"o}nigsberg's bridges have been demolished since Euler's time, the resulting theory and techniques continued to flourish with applications increasing both in number and scope.}

\textcolor{black}{The first textbook on graph theory \cite{konig1990theory} appeared in 1936, i.e., nearly two centuries after its first application by Euler. Nowadays, graph theory is considered a \emph{fundamental} tool with numerous available references. For example, the authors in \cite{west2001introduction} present a thorough introduction to graph theory, its concepts, and algorithms. Algorithms for several other graph theory problems, e.g., path searching, tree counting, planarity, can be found in the following reference \cite{gould2012graph}. The textbooks by Bollob{\'a}s \cite{bollobas2013modern} and Diestel \cite{diestel2000graph} provide an in-depth investigation of modern graph theory tools including flows and connectivity, the coloring problem, random graphs, and trees.}

\textcolor{black}{Current applications of graph theory span several fields. Indeed, graph theory, as a part of discrete mathematics, is particularly helpful in solving discrete equations with a well-defined structure. Reference \cite{ba3na2002walk} provides multiple connections between graph theory problems and their combinatoric counterparts. Multiple tutorials on the applications of graph theory techniques to various problems are available in the literature, e.g., \cite{vasudev2006graph,4389477,6334718,8389737,7046827,4375569,918502,6005872}. In particular, while the authors in \cite{vasudev2006graph} provide a general introduction to graph theory with its traditional applications, reference \cite{4389477} focuses on the usefulness of spectral graph theory \cite{chung1997spectral} for clustering and image segmentation. Several specific applications of graph theory are available such as system recovery \cite{6334718}, image segmentation \cite{8389737}, bio-engineering \cite{7046827,4375569}, power systems \cite{918502}, and computer science \cite{6005872}. Unlike all aforementioned works, the current tutorial focuses on a particular problem in graph theory, known as the clique problem, and its applications in solving optimization and design problems in communications and signal processing. The manuscript illustrates the structural properties of programs that can be formulated and solved as clique problems through recent communications and signal processing applications. To the best of the authors' knowledge, this is the first tutorial that focuses on the clique problem, its variants, and their modern applications in communications and signal processing problems.}

\textcolor{black}{The first part of the manuscript provides the fundamental definitions and tools of graph theory. In particular, the maximum clique, maximum weight clique, and $k$-clique problems are formulated, and multiple optimal and heuristic solutions are presented. The most relevant benefit of casting a problem into a clique problem, whenever possible, is that it can be solved reliably and efficiently by exploiting the diverse set of numerical solutions available in the literature. Towards that end, the tutorial regards the clique problem as an integer program whose structural properties are identified, and for which integer solutions are provided. These integer methods are particularly interesting for problems wherein additional information about the problem can be exploited and/or the construction of the graph is either not feasible or excessively complicated. The tutorial, further, illustrates the use of the clique formulation and its variants through numerous contemporary examples in communications and signal processing, mainly in maximum access for non-orthogonal multiple access networks, throughput maximization using index and instantly decodable network coding, collision-free radio frequency identification networks, and resource allocation in cloud-radio access networks. Finally, the tutorial sheds light on the recent advances of such applications, and provides technical insights on ways of dealing with mixed discrete-continuous optimization problems.}

The remaining of this tutorial is divided as follows. \sref{sec:gra} introduces the clique problem and its variants and suggests optimal and heuristic solutions. In \sref{sec:opt}, the clique problem is regarded and solved as an integer optimization problem. Sections \ref{sec:gra2}, \ref{sec:gra3}, and \ref{sec:gra4} illustrate the formulation of contemporary problems in communications and signal processing as maximum clique, maximum weight clique, and $k$-clique problems, respectively. Finally, before concluding in \sref{sec:con}, \sref{sec:gra5} presents recent advances of the proposed applications, and sheds light on ways of dealing with mixed discrete-continuous optimization.

\section{The Clique Problem in Graph Theory: Definitions, Formulation and Algorithms} \label{sec:gra}

\textcolor{black}{This section introduces the clique problem and its variants from a graph theory perspective, and suggests optimal and heuristic solutions for finding cliques. In particular, the section focuses on providing notations and definitions in graph theory, and on formulating the clique problems. Afterward, multiple graph-based exact and efficient algorithms for finding cliques are presented, and their computational complexities are investigated. Finally, the section presents specialized clique related algorithms for graphs with well-defined structures.}

\subsection{Definitions and Problem Formulation}

\subsubsection{Notations and Definitions}

\textcolor{black}{A graph $\mathcal{G}(\mathcal{V},\mathcal{E})$ is a collection of two \emph{finite} sets $\mathcal{V}$ and $\mathcal{E}$ that can be naturally expressed by a visual representation. In such illustrations, vertices are usually denoted by circles and edges by lines (or arrows in the case of directed graphs). The set $\mathcal{V}$ is called the set of vertices of the graph, and a vertex $v \in \mathcal{V}$ is referred to as a node in the graph. The cardinality of $\mathcal{V}$, denoted by $n=|\mathcal{V}|$, represents the total number of vertices, i.e., the size of the graph. The set of edges $\mathcal{E}$ represents the ensemble of connections between pairs of vertices. Such a set contains either ordered or non-ordered pairs of vertices, which gives rise to the notion of direction in graphs. The study of directed graphs is crucial for applications in which the direction of the flow matters, e.g., causal structures and time-varying systems, and is particularly interesting for communications and signal processing applications, as shown in \sref{sec:gra5}. However, due to the predominance of undirected graphs for the applications of interest herein, directed graphs are discussed briefly as structured graphs later in the section.}

\textcolor{black}{Besides the above-mentioned notion of direction, graphs can be classified according to their weights. In particular, a graph is said to be vertices (edges, respectively) weighted if its vertices (edges, respectively) have weights. Applications of edges weighted graphs include the notorious shortest path problem \cite{schrijver2012history}, which belongs to a few graph theory problems with a polynomial-time solution, e.g., the Dijkstra's algorithm \cite{dijkstra1959note}. Despite their importance, edges weighted graphs are omitted herein as vertices weighted graphs are more generic. Indeed, edges weighted graphs can be transformed into vertices weighted graphs by the introduction of intermediate nodes. Therefore, only vertices weighted graphs, called simply weighted graphs, are considered in the remainder of this tutorial. Furthermore, throughout the manuscript, the weight of a vertex $v \in \mathcal{V}$ is denoted by $\omega(v) \in \mathbb{R}$ and is assumed to be positive. Indeed, clique problems are agnostic to negative weights, for they can be removed without altering the solution. Nevertheless, the study of such graphs is relevant to several applications that fall outside the scope of this manuscript.}

\begin{figure}[t]
\centering
\includegraphics[width=0.6\linewidth]{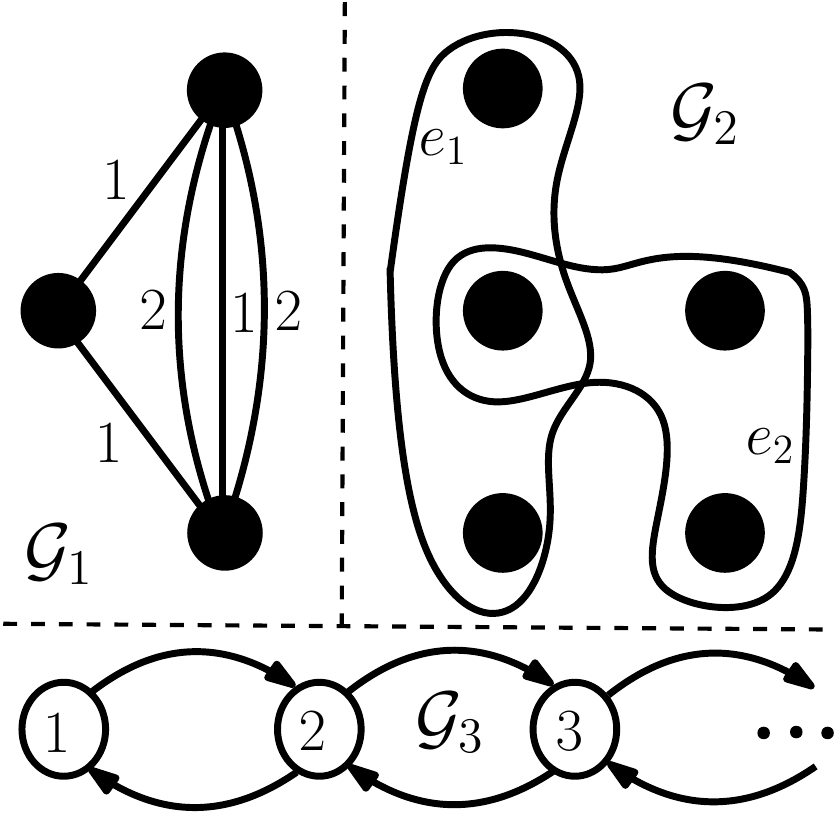}\\
\caption{Example of a multigraph, a hypergraph, and an infinite graph. The top left corner represents a weighted multigraph. The top right corner illustrates a hypergraph with two hyper-edges $e_1$ and $e_2$. The bottom represents a directed infinite graph in which both the set of vertices and the set of edges are infinite.}\label{fig:4}
\end{figure}

It is worth mentioning that the definition of graphs given herein can be extended in multiple directions. In addition to the aforementioned notion of direction and weight, various other variants of graphs exist. For example, graphs with a multi-set of edges $\mathcal{E}$, i.e., the same connection can exist multiple times, are called multigraphs. These graphs are especially attractive for the study of resilience in networks. Indeed, pieces of equipment can be linked with multiple physical connections to provide extra resilience against link failures. Similarly, allowing edges to connect an arbitrary number (rather than just a pair) of vertices results in a hypergraph. Finally, the study of infinite graphs, in which either $\mathcal{V}$ and/or $\mathcal{E}$ are infinite, is particularly interesting for unveiling the properties of Markov chains. \fref{fig:4} illustrates a multigraph, a hypergraph, and an infinite graph. While these generalizations of graphs are interesting, their applications are generally out of the scope of the current tutorial.

\subsubsection{The Clique Problem Formulation}

\textcolor{black}{Given a graph $\mathcal{G}(\mathcal{V},\mathcal{E})$, a clique is a subset of vertices such that each vertex in the set is adjacent to all other vertices in the set. In other words, a subset $\mathcal{C} \subseteq \mathcal{V}$ is a clique if and only if $(v,v^\prime) \in \mathcal{E}$ for all vertices $v$ and $v^\prime \in \mathcal{C}$. Cliques often unveil the tangible properties of graphs. For example, assuming that nodes in the graph represent people and edges friend relationships, then a clique is a set of individuals who are all friends with each other. Indeed, the word \emph{clique}, borrowed from French and first suggested in a 1949's study of the above social network model by Luce and Perry \cite{luce1949method}, refers to a group of people with a common interest. Given the notion of cliques, a maximal clique in a graph is a clique that cannot be extended by including more adjacent vertices without compromising its connectivity property, i.e., it is not a subset of a larger clique. Similarly, the maximum clique is the maximal clique with the highest number of vertices.}

\textcolor{black}{While cliques exhibit local properties of the graph, the maximum clique reveals global trends and plays a significant role in communications and signal processing applications. For example, considering the aforementioned social network model, one can readily see that the maximum clique is arguably one of the most intuitive approaches for clustering, i.e., grouping people into communities. The maximum clique problem refers to the problem of finding the maximum clique in a given graph. Similarly, the maximum weight clique problem is the problem of finding the maximal clique with the maximum weight. On the other hand, the $k$-clique problem adds the extra constraint that the size of the clique should be at least $k$ for some positive integer $k \geq 1$. Likewise, the maximum weight $k$-clique problem is the one of finding the clique with the maximum weight and of size at least $k$. Mathematically, the maximum weight $k$-clique problem can be formulated in the graph $\mathcal{G}(\mathcal{V},\mathcal{E})$ as}
\begin{subequations}
\label{cliqueorginal1}
\begin{align}
\max_{\mathcal{C} \subseteq \mathcal{V}} \ & \sum_{v \in \mathcal{C}} \omega(v) \\
 {\rm s.t.\ }  & (v,v^\prime) \in \mathcal{E},\ \forall \ v,v^\prime \in \mathcal{C}, \\
& \sum_{v \in \mathcal{C}} 1 = |\mathcal{C}| \geq k.
\end{align}
\end{subequations}

The maximum weight and $k$-clique problems can be obtained as particular instances of the problem in \eref{cliqueorginal1} wherein the size $k$ and the weights $\omega(v)$ are set to $1$, respectively. Finally, the maximum clique problem can be obtained by setting $\omega(v) = k = 1$ in the above formulation. The graph-based problem formulation in \eref{cliqueorginal1} is revisited in \sref{sec:opt} to provide alternative constructions of the clique problem from an integer programming perspective, which is particularly useful for recognizing problems that are solvable by a clique search.

\textcolor{black}{Unlike cliques, an independent set is a subset of vertices such that no connection exists between any pair of vertices in the set. It can easily be seen that the clique and independent set problems are complementary by considering the complement of the graph, i.e., inverting all edges. Finally, in addition to the concept of cliques introduced in this section, it is worth mentioning the existence of a noisy version of a clique called a paraclique \cite{hagan2016lower}. For applications in which graphs are constructed from noisy experimental data, the concept of cliques is too restrictive. The paraclique connotation relaxes the definition of a clique to account for noise. Given a clique, the paraclique includes all vertices with a predefined proportion of edges from the chosen core clique. This clustering approach is shown to be particularly well-adapted for biological data clustering, e.g., \cite{chesler2007combinatorial,rokas2003genome}, but falls outside the scope of this tutorial.}

\subsection{Graph-Based Clique Algorithms and Complexity} \label{sec:subqwe}

\textcolor{black}{This subsection presents multiple optimal and heuristic graph-based clique algorithms, together with their computational complexities. In particular, the first part suggests optimal algorithms for clique search and classifies them according to their clique generation method. Afterward, the complexity of these methods is investigated, which motivates the need for presenting heuristics capable of producing satisfactory solutions in polynomial time.}

\subsubsection{Optimal Solutions to the Clique Problem}

\textcolor{black}{The clique problem is one of the most studied problems in graph theory, as indicated by the sheer number of proposed exact solutions with varying requirements on the size of the graph, its structure, size of cliques, storage memory, etc. Although different, most of these algorithms fall under one (or more) of the following approaches according to the method used to generate cliques:}
\begin{itemize}
\item Vertex removal,
\item Backtracking (depth-first search),
\item Breadth-first search,
\item Algebraic decomposition,
\item Integer programming.
\end{itemize}

\textcolor{black}{The vertex removal approach relies on generating cliques of a graph $\mathcal{G}(\mathcal{V},\mathcal{E})$ from the cliques of an included graph $\mathcal{G}-\{v\}$, for some vertex $v \in \mathcal{V}$. In other words, the method gradually includes vertices in the discovered cliques while maintaining the list of all previously found smaller cliques, e.g., see \cite{augustson1970analysis,osteen1973clique,osteen1974clique}. This process generally results in finding small maximal cliques early. However, one major drawback of this method is that it requires a large memory to store all previously discovered cliques. Indeed, in order to ensure that the discovered cliques are maximal, each new clique needs to be compared with all cliques listed in previous steps. While the approach has been extensively used to derive clique search algorithms \cite{tsukiyama1977new,lawler1980generating,chiba1985arboricity}, it is not suitable for solving the $k$-clique problem, as small cliques are discovered first.}

\textcolor{black}{The backtracking approach generates a tree of possible solutions, i.e., possible cliques, and systematically traverses the tree using a depth-first search routine. The method is particularly interesting as it avoids the generation of duplicated cliques, which alleviates the need to store all previous solutions to check for maximality. Therefore, while the vertex removal approach requires an exponential storage size \cite{moon1965cliques}, the backtracking method requires only polynomial storage space and is extremely popular for practical implementations, e.g., \cite{akkoyunlu1973enumeration,johnston1976cliques,gerhards1979clique,loukakis1981depth,loukakis1983new,johnson1988generating,koch2001enumerating,eppstein2001small,harley2001uniform,harley2003graph,eppstein2003small,tomita2004worst,tomita2006worst,wan2006new,regneri2007finding}. In backtracking, however, there is no particular order in which the cliques are discovered, and numerical experimentation suggests that cliques of different sizes are discovered in a quasi-random fashion. One significant advantage of the backtracking method, on the other side, is that it avoids the exploration of some branches of the search tree which do not lead to new cliques.}

\textcolor{black}{Similar to the backtracking approach, the breadth-first search algorithm, as its name indicates, systematically traverses the tree of possible solutions using a breadth-first search routine. A notable advantage of the breadth-first search over the depth-first search is that it enumerates maximal cliques in non-decreasing size order \cite{kose2001visualizing}. It shares, however, the same storage size drawback of the vertex removal, as all cliques need to be maintained in memory to check for maximality.}

\textcolor{black}{The algebraic approach treats the graph from a linear algebraic perspective, and applies standard tools to discover the largest clique. For example, algebraic decomposition methods seek to decompose the graph of interest into a chain of subgraphs, each containing fewer cliques than their union and to employ a quasi-block diagonalization. This approach is notably exploited in \cite{makino2004new} to solve the maximum clique problem.}

\textcolor{black}{Finally, the integer programming approach suggests regarding the clique problem as an integer program and aims at solving it using generic integer program techniques, e.g., the branch-and-bound (BnB) algorithm. In general, such approaches lead to poorer performance than using algorithms specifically designed for cliques. Nevertheless, for applications in which additional information about the problem can be exploited and/or the construction of the graph is either not feasible or excessively complicated, integer programming methods are a clear winner against graph-based algorithms. Given their richness and importance, the presentation of integer programming techniques for the clique problem is presented in details later in next section.}

\subsubsection{Complexity of Optimal Clique Search Methods}

Although easy to state, the maximum clique problem, and by extension, its variants are difficult to solve. Graph-theoretical approaches, especially those that enumerate all maximal cliques, are often memory-intensive and must share huge sets of data efficiently across many processors in parallel implementations as a graph with $n$ nodes can have as many as $3^{n/3}$ maximal cliques \cite{moon1965cliques}. Furthermore, the memory requirement grows exponentially with the size of the graph and can reach a terabyte-scale for modern applications. Indeed, the clique problem is listed as one of Karp's 21 NP-complete problems \cite{karp1972reducibility}. Furthermore, the problem is intractable and hard to approximate \cite{garey2002computers}. The above concepts in complexity theory are not further detailed in the text, and interested readers are referred to books \cite{harris2008combinatorics,ausiello2012complexity} in which the authors provide a comprehensive investigation of the approximation complexity and the relationship between combinatorics and graph theory problems through multiple examples.

\textcolor{black}{Due to its NP-hardness, exact solutions exhibit an exponential complexity behavior. In other words, for a graph $\mathcal{G}$ with $n$ nodes, the complexity of optimal algorithms scales as $\mathcal{O}(\alpha^{n})$, for some parameter $1 < \alpha \leq 2$. For example, a brute force method would list and check all possible cliques resulting in a total complexity of $n^2 2^{n} = \mathcal{O}(2^{n})$. A clever use of the particular structure of the problem makes it possible to reduce the complexity of the solution from $2^n$ to $\alpha^n$, where $1 < \alpha < 2$, e.g., see the algorithms in \cite{tarjan1977finding,jian19862,robson1986algorithms,15522856,2155446}, the complexity of which is summarized in \tref{tab:1}. Note that the complexity is obtained by extensive simulations and averaged over a large number of graphs with different properties. Most of the above-cited algorithms are further extended to solve the maximum weight clique and the $k$-clique problems, e.g., the maximum weight clique search in \cite{13265492}. Interested readers are referred to \cite{johnston1976cliques,pardalos1994maximum,bomze1999maximum} for a more comprehensive survey on traditional clique algorithms and their complexities.}

\begin{table}
\caption{Complexity of Optimal Maximum Clique Algorithms.}
\centering
\begin{tabular}{|c|c|}
\hline
Algorithm  & Complexity\\
\hline
Tarjan and Trojanowski \cite{tarjan1977finding} & $1.261^{n}$\\
Jian \cite{jian19862} & $1.235^{n}$\\
Robson \cite{robson1986algorithms} & $1.211^{n}$\\
Fomin et al. \cite{15522856}& $1.221^{n}$\\
Bourgeois et al. \cite{2155446}& $1.212^{n}$\\
\hline
\end{tabular}
\label{tab:1}
\end{table}

\subsubsection{Efficient Heuristics to Discovering Cliques} 

\textcolor{black}{All above mentioned exact algorithms have a worst-case exponential running time in the size of the graph. In practice, however, obtaining the optimal solution may not be desirable for computation and energy consumption considerations. In such scenarios, one is interested in finding a ``good enough" maximal clique in a reasonable time, which is achieved using advanced heuristic algorithms. As such, various efficient heuristic algorithms have been developed for the different variants of the clique problem, e.g., see \cite{9874286,bomze1999maximum} for a summary of multiple solutions. These greedy algorithms rely on various routines to solve the problem in polynomial time. For example, the authors in \cite{babel1990branch} present a BnB-based greedy algorithm for the maximum clique problem that explores the branches containing vertices with high connectivity. It is observed that the heuristic discovers a maximum clique sooner than exact algorithms, but takes significantly longer to verify optimality.}

References \cite{7103958} and \cite{6607889} derive polynomial-time solutions to the maximal clique problem using a recursive approach. Unlike optimal solutions that systematically explore all branches of the solution tree, these heuristics develop a pruning strategy to abort the exploration of small size cliques. The authors in \cite{6848102} design maximum weight clique heuristics for wireless networking applications. In particular, for arbitrary interference radii, the authors suggest finding constant-approximate solutions efficiently. Similarly, the authors in \cite{5341909} solve the maximum weight clique problem in polynomial time for a specific class of graphs.

The design of efficient heuristics for the maximum $k$-clique problem is of great interest for clustering and community detection applications, in which the large size of the graph makes optimal solutions of less importance. In \cite{7117352}, an efficient algorithm for the maximum $k$-clique problem is designed using a modified adjacency matrix and formal concept analysis, i.e., a typical computational intelligence technique. Similarly, the authors in \cite{7117352} suggest solving the $k$-clique community detection in social networks using a union-find structure to store divided communities and reduce the number of unnecessary intersection test. Extensive simulations on real data reveal that the algorithm is reasonable, effective, and divides all communities within approximately linear time complexity.

For illustration purposes, the manuscript explains the outlines of a simple quadratic complexity greedy algorithm to find the maximum weight clique in a graph. The algorithm relies on modifying the weights of each vertex by the average weight of its neighbors. This step assigns high weights to vertices that have a large initial weight and are connected to multiple nodes with high weights. Afterward, the vertex with the highest modified weight is selected and removed from the graph. The graph is updated to keep only nodes that are connected to the removed vertex. Therefore, the final selected set represents a clique in the graph. The steps of the algorithm are illustrated in \algref{alg2}.

\begin{algorithm}[t!]
\begin{algorithmic}[1]
\REQUIRE Graph $\mathcal{G}(\mathcal{V},\mathcal{E})$, weights $w(v)$, and neighbors $N(v)$.
\WHILE {$\mathcal{G} \neq \varnothing$}
\FORALL{$v \in \mathcal{V}$}
\STATE $w^\prime(v) \leftarrow \sum_{v^\prime \in N(v)} w(v^\prime)$.
\ENDFOR
\STATE $v_i \leftarrow \arg\max_{v \in \mathcal{V}} (w^\prime(v))$.
\STATE $C \leftarrow C \cup \{v_i\}$.
\STATE $\mathcal{G} \leftarrow  N(v_i)$.
\ENDWHILE
\RETURN $C$.
\end{algorithmic}
\caption{A Simple Maximum Weight Clique Heuristic}
\label{alg2}
\end{algorithm}

The complexity of \algref{alg2} is controlled by the weight modification and vertex selection procedures. While the weight modification procedure is quadratic in the number of vertices, the vertex selection process is linear. Therefore, the overall complexity of the algorithm is quadratic in the size of the graph. Note that the concept of complexity in this section refers to the complexity of solving the clique problem and its variants. This does not include the complexity of constructing the graph. Such complexity eventually plays a vital role in the overall efficiency of the algorithms proposed to solve communications and signal processing problems, as discussed in the \sref{sec:opt}.

\subsection{Specialized Algorithms for the Clique Problem}

As its name indicates, this subsection provides some specialized clique algorithms relying on novel and unique approaches, e.g., quantum-inspired evolutionary algorithms. While the first part of the subsection cites some particular clique algorithms for structured graphs, e.g., sparse, irregular, and large graphs, the second part surveys novel and unique approaches for the clique problem and its variants that are inspired by the advances in other scientific disciplines, e.g., quantum computing, biology, and so on. Afterward, the tutorial emphasizes the role of directed graphs in representing time-varying systems and suggests communications and signal processing applications of the clique problem in directed graphs.

\subsubsection{Specialized Clique Algorithms for Structured Graphs}

\textcolor{black}{As stated earlier, the maximum clique problem, its variants, and even their approximations are often hard to solve \cite{garey2002computers}. In many practical settings, however, graphs may have particular structures, e.g., sparse and irregular graphs, which makes finding the maximum clique feasible in a reasonable time using advanced optimal and heuristic algorithms. Indeed, a large body of literature has been dedicated to solving the maximum clique problem for various structured graphs.}

The most studied, relevant, and common particular structures for graphs are unarguably sparse and large graphs. Indeed, for some applications, say social networks, graphs are so massive, and probably also sparse, such that traditional single computer solutions are no longer effective and face significant challenges. These graphs receive considerable attention in the literature thanks to their wide use in numerous engineering applications. For example, reference \cite{pardalos1999maximum} suggests discovering the maximum clique in very large graphs by breaking the problem in several pieces of manageable size. The author in \cite{eppstein2009all} proposes a depth-first search based algorithm to solve the maximum independent set in large graphs. The algorithm is revised in \cite{eppstein2010listing} to list all maximal cliques in sparse graphs in near-optimal time. It is worth noting that solving the maximum clique in large graphs is still an active area of research, e.g., see \cite{31376280,8488050} for new algorithms to find, respectively enumerate, maximal cliques in large-scale networks.

\textcolor{black}{Besides breaking the problem into smaller sizes, a large body of literature has been dedicated to proposing distributed algorithms that can be executed concurrently in a computer cluster. For example, the authors in \cite{carraghan1990parallel,pardalos1998exact,bomze2000approximating,6249683} consider large graphs and present parallel algorithms that take advantage of multiple parallel processing units to solve the clique problem. In particular, reference \cite{carraghan1990parallel} introduces a parallel greedy algorithm for the maximum weight clique problem which is improved in \cite{bomze2000approximating} using replicator dynamics. An exact parallel algorithm to solve the maximum clique problem is designed in \cite{pardalos1998exact}. The authors in \cite{6249683} consider the $k$-clique problem variant for a community detection application in large-scale social networks for which a parallel $k$-clique algorithm is proposed with theoretical tight upper bounds on its worst-case time and space complexities. Nonetheless, note that solving the maximum clique distributively is still an active area of research. For example, recently, reference \cite{8540236} proposed encoding the maximum clique problem as a partial maximum satisfiability program, and adapted the Dist algorithm \cite{cai2014tailoring} to solve it.}

\subsubsection{Novel Approaches for the Clique Problem}

The maximum clique problem and its variants are receiving considerable and continuous interest from the research community for their numerous and practically universal applications, e.g., the new paradigm of data caching \cite{8265823}. Besides the classical approaches identified above, the recent years witnessed a surge of clique algorithms inspired by the advances in other fields, e.g., quantum computing, biology, etc. For instance, reference \cite{8394892} provides a good quality solution to the maximum weight clique problem in reasonable computational times. This is accomplished by casting the maximum weight clique problem as an integer program and designing a local search method that avoids entrapment in local optima thanks to a couple of perturbations. Likewise, the authors in \cite{bonamy2018eptas} propose a polynomial-time algorithm so as to discover the maximum clique in unit ball graphs.

Further, several recent works get inspiration from quantum computing \cite{hey1999quantum,kaye2007introduction} and biology to design novel graph theory algorithms, e.g., see \cite{han2002quantum,das2014effective,das2014quantum} and references therein. In particular, the authors of \cite{7307535} use quantum concepts to design novel maximum clique algorithms. Reference \cite{7307535}, particularly, proposes a quantum-inspired evolutionary algorithm to solve the maximum clique problem by using a population of Q-bit individuals and Q-gate as the main variation operators. The first effective implementation of the maximum independent set routine on a quantum computer can be found in \cite{8477865}. On the other hand, the authors in \cite{5338147} suggest a molecular beacon-based DNA computing model for solving the maximum weight clique problem. The authors in \cite{5338147} propose encoding the vertices weights into unique fixed-length oligonucleotide segments and solving the problem using a sticker model. Unlike traditional approaches, the solution of \cite{5338147} is memory efficient, as it does not require generating an initial data pool that contains every possible solution to the problem of interest.

\textcolor{black}{Structured graphs, and in particular large graphs, also benefited from the aforementioned novel approaches. For example, while reference \cite{funabiki1992neural} suggests a neural network model for finding a near-maximum clique, the authors in \cite{funabiki1992neural} design a distributed algorithm for the clique problem with good convergence rate and scalability enabled by the proposed pruning routine to reduce the solution space. The algorithm is extended in \cite{6544815} and \cite{6906774} to provide fault-tolerance for a quantum error-correcting codes application and a reduced worst-case time complexity for computing maximal cliques in common real-world graphs, respectively. Unlike the aforementioned synchronous communication distributed algorithms, reference \cite{8287735} suggests asynchronous distributed-memory parallel maximum independent set algorithms based on a directed acyclic graph.}

\subsubsection{Time-Varying Systems as Directed Graphs}

Graph theory plays a fundamental role in representing data and complex networks with broad applications in several areas. While all previously mentioned algorithms are designed for undirected graphs, some recent applications induce highly structured graphs for which specialized algorithms are expected to outperform classical solutions. A particularly interesting type of structured graphs is time-varying graphs (TVG) for their ability in modeling time-varying and dynamic complex networked systems \cite{casteigts2011time,sengul2012survey}. These graph representations are studied in the literature to model dynamic processes over complex networks, i.e., random walks or information diffusion. For example, in \cite{8449969}, a two-stage stochastic optimization approach for finding cliques in randomly changing graphs is proposed.

In time-varying graphs, the graph structure, i.e., the set of nodes and edges, evolves as time passes. The use of classical graph theory algorithms in TVG may not be feasible as these algorithms are not designed to consider the time relations between nodes. There has been a good number of attempts to extend classical graphs into TVG, leading to different models known in the literature as time-dependent networks \cite{casteigts2012time,alvarez2012complex,holme2012temporal}. For example, the authors in \cite{viard2016computing} consider a particular TVG known as link stream and extend the notion of a clique to a $\Delta$-clique for some time interval $\Delta$, such that all pairs of nodes in this $\Delta$-clique interact at least once during each sub-interval of duration $\Delta$. A maximum $\Delta$-clique enumeration is designed and evaluated using real-world data. The aforementioned framework is exploited in \cite{viard2018discovering} for discovering patterns of interest in IP traffic using cliques in bipartite link streams.

\textcolor{black}{The authors in \cite{7344810} provide a unified and general framework for representing finite and discrete time-varying graphs by directed graphs. The proposed framework extends the results of \cite{wehmuth2z016multiaspect,wehmuth2016multiaspect} and can represent several previous models for dynamic networks found in the recent literature. By demonstrating that multi-layer and time-varying networks are isomorphic to directed static graphs, the directed graph model of \cite{7344810} preserves the strictly discrete nature of the basic graph abstraction, while allowing to represent time relations between nodes properly. Therefore, given the importance of directed graphs in representing time-varying systems, e.g., \cite{8495644}, there has been a surge in the recent literature to adapt classical graph theory algorithms to directed graphs. For example, reference \cite{7035970} designs an independent set search routine for directed graphs by exploiting the quantum computing paradigm. The resulting algorithm exhibits an almost linear complexity for most cases. Likewise, the authors in \cite{meeusen1975clique} suggest an algorithm for clique detection in directed graphs; reference \cite{conte2017maximal} provides a tight bound on the number of cliques in a directed graph, and highlights their useful structural properties.}

\section{An Optimization Approach for Finding Cliques} \label{sec:opt}

\textcolor{black}{The previous section considers the clique problem from a graph theoretical perspective. In many applications, however, it is beneficial to formulate the clique problem as an integer program for scenarios wherein additional information about the problem can be exploited and/or the construction of the graph is either not feasible or excessively complicated. In such situations, it becomes more advantageous to solve the clique problem by using integer programming techniques. Hence, this section regards the clique problem from an algebraic and optimization perspectives. Indeed, in addition to the standard definition of a graph given in \sref{sec:gra}, a graph can be defined by its adjacency matrix, usually denoted by $\mathbf{A}$. The adjacency matrix is a $n \times n$ $0-1$ matrix, wherein $n$ represents the number of vertices in the graph. Denoting a graph by its adjacency matrix allows the use of classical linear algebra tools to analyze the properties and structure of the graph. For example, the connectivity of the graph can be attested by the strict positiveness of the second eigenvalue $\lambda_2 >0$, known as the algebraic connectivity \cite{458200753}, of the Laplacian matrix $\mathbf{L}$ of the graph \cite{25181258}. Such graph metrics are particularly interesting for applications requiring the connectivity of the graph, e.g., backhaul network planning \cite{7511143,7306543,5462107,7458183}. For instance, references \cite{7511143} and \cite{7247067} exploit the clique formulation to solve the problem of minimizing the cost of backhaul design under connectivity, data rate, reliability, and resilience constraints. Furthermore, the approach allows transforming the graph theory problems to generic integer programs, which can be beneficial under some conditions, as shown later in the section.}

\textcolor{black}{This section first formulates the clique search as a linear integer program, which allows identifying the structural properties of the problem. The section proposes optimal generic integer algorithms for solving the integer program, including the cutting planes method and the branch-and-bound algorithm. Further, the section proposes meta-heuristics and evolutionary algorithms for their universality and fixed complexity features.}

\subsection{Identifying Clique Problems using Integer Linear Programs}

The maximum clique problem has various integer linear program formulations such as the edge formulation, the independent set formulation, and the quadratically constrained global optimization problem. A survey of the different formulations of the clique problem as an integer program and its various relaxations along with the appropriate algorithms is available in \cite{Bomze99themaximum}. Showing that a problem is equivalent to a clique problem (see \sref{sec:gra2}, \sref{sec:gra3}, and \sref{sec:gra4}) eventually reduces to demonstrating that its formulation matches one of the clique integer program formulations. This section focuses on a practical formulation of the maximum clique problem, known as the edge formulation \cite{nemhauser1975vertex}, to identify problems that can be solved by a clique-finding routine.

Let $\mathcal{G}(\mathcal{V},\mathcal{E})$ be the graph of interest containing $n$ nodes. The maximum clique problem can be formulated as:
\begin{subequations}
\label{clique1}
\begin{align}
\max_{x_{i} \in \{0,1\}} & \sum_{i=1}^n \omega_i x_i \\
 {\rm s.t.\ }  & x_{i} + x_{j} \leq 1,\ \forall \ (i,j) \notin \mathcal{E},
\end{align}
\end{subequations}
where the optimization is over the binary variables $x_i$'s. The weights $\omega_i$ are set to $1$ for the maximum clique problem, and to the weight of the corresponding vertex for the maximum weight clique problem. The $k$-clique problem can be obtained from the above formulation by constraining the size of the found clique to be at least $k$. In other words, the $k$-clique problem can be formulated as follows:
\begin{subequations}
\label{clique2}
\begin{align}
\max_{x_{i} \in \{0,1\}} & \sum_{i=1}^n \omega_i x_i \\
 {\rm s.t.\ }  & x_{i} + x_{j} \leq 1,\ \forall \ (i,j) \notin \mathcal{E} \\
& \sum_{i=1}^n x_i \geq k,
\end{align}
\end{subequations}
Therefore, in order to figure whether a generic discrete optimization problem can be solved by a clique search, one needs to check whether it is possible to recast the discrete optimization problem in a form equivalent to \eref{clique2}. In other words, consider a generic discrete optimization problem with a \emph{linear} objective function and \emph{linear} equality constraints:
\begin{subequations}
\label{clique3}
\begin{align}
\max_{\mathbf{x} \in \{0,1\}^n} &\ \mathbf{c}^T\mathbf{x} \\
 {\rm s.t.\ }  & f_i(\mathbf{x}) \leq 0,\ 1 \leq i \leq m \\
& g_j(\mathbf{x}) = 0,\ 1 \leq j \leq k.
\end{align}
\end{subequations}

Problem \eref{clique2} is equivalent to a clique problem if and only if one can design a set $\mathcal{E} \subseteq \mathbb{R}^n \times \mathbb{R}^n$ such that for all indices $(s,t) \notin \mathcal{E}$ and for all feasible solution $\mathbf{x} \in \mathbb{R}^n$ of \eref{clique3}, i.e., $f_i(\mathbf{x})\leq 0,\ 1 \leq j \leq m$ and $g_j(\mathbf{x}) = 0,\ 1 \leq j \leq k$, the vector $\mathbf{x}$ satisfies: $\mathbf{x}_s \mathbf{x}_t =0$, where $\mathbf{x}_s$ and $\mathbf{x}_t$ are the $s$-th and $t$-th entries of the vector $\mathbf{x}$. The above designed set $\mathcal{E}$ corresponds to the set of edges in the graph $\mathcal{G}(\mathcal{V},\mathcal{E})$, where the set of vertices is given by $\mathcal{V} = \{\mathbf{x}_{i}\}_{i=1}^n$.

The design of the set of connections $\mathcal{E}$ depends on the structure of the problem and should be constructed on a case by case basis. Given the connectivity constraints, a popular approach to show the equivalence of both problems consists of finding a one-to-one mapping between all feasible solutions and all cliques in the graph. Sections \ref{sec:gra2}, \ref{sec:gra3}, and \ref{sec:gra4} illustrate how to construct such set of connections for various communications and signal processing applications.

\subsection{Exact Integer Program Methods}

\textcolor{black}{Generic integer program methods can be classified into two main categories of algorithms, viz.:
\begin{itemize}
\item Exact algorithms that are guaranteed to find an optimal solution. These algorithms, however, have exponential worst-case complexity. The tutorial focuses on two such algorithms, namely, the cutting planes method \cite{marchand2002cutting}, and the branch-and-bound algorithm \cite{pardalos1992branch}. Both of these algorithms rely on two basic concepts, namely relaxations and bounding.
\item Greedy algorithms that provide a sub-optimal solution without any guarantee on its optimality. However, these algorithms usually have a low or a fixed complexity and usually rely on a stochastic process to increase their probability of finding a good solution. These approaches are provided in the next subsection.
\end{itemize}}

\subsubsection{The Cutting Planes Method}

The cutting planes method is a class of iterative algorithms that have been first proposed by Ralph E. Gomory to solve linear integer programs by iteratively refining the search space \cite{marchand2002cutting}. The fundamental idea behind cutting planes is to add linear constraints to a linear program until the optimal basic feasible solution takes on integer values.

Such a method solves the integer problem optimally, i.e., the added constraints help to reach the optimal solution. More precisely, the method drops the requirement that the variables are integers and solves the associated linear programming problem to obtain a basic feasible solution. If the found solution is not an integer, the method finds a hyperplane with the vertex on one side and all feasible integer points on the other. In other words, it adds a linear constraint, also called a cut, to produce a half-space such that the current found solution lies on one side of the half-space, while every feasible integer solution lies on the other side. The same technique is applied to the new program and repeated until an integer solution is discovered.

The performance of the cutting planes method largely depends on the choice of cuts. The first version of the algorithm in the 1950s was deemed inefficient, as it requires many rounds of cuts to make progress towards the solution. Indeed, many cuts are either expensive or even NP-hard to reach, which results in numerical instability. A combination of the cutting planes and the branch-and-bound algorithm, known as branch-and-cut \cite{padberg1991branch}, may result in a much more stable and viable algorithm for which there are two ways to generate cuts \cite{cornuejols2008valid}.

\subsubsection{The Branch-and-Bound Algorithm}

\begin{figure}[t]
\centering
  \includegraphics[width=0.4\linewidth]{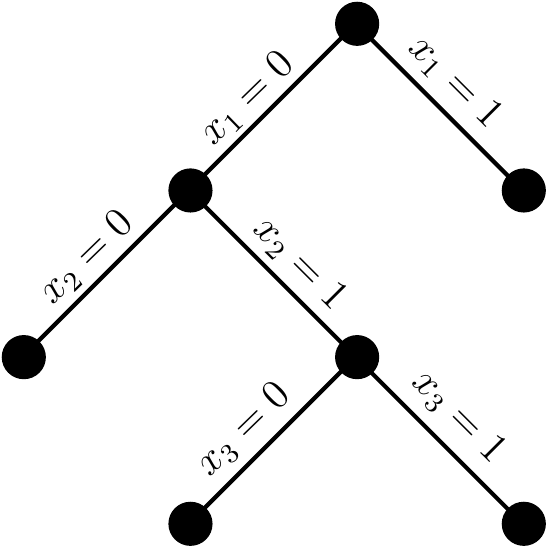}\\
  \caption{Branch and bound for a problem with $3$ discrete variables $x_1,x_2$, and $x_3$. The branch for $x_1=1$ is not explored as the optimal solution is provable not in that subspace. The same goes for the branch $x_1=0,x_2=0$.}\label{fig:branch}
\end{figure}

\textcolor{black}{The branch-and-bound algorithm is first introduced in \cite{land1960automatic} for solving discrete, i.e., integer, programs. The algorithm is popularized with its successful implementation by Little et al. \cite{little1963algorithm} for solving the traveling salesman problem, a famous proven NP-hard problem. BnB relies on a systematic enumeration of candidate solutions similar to an exhaustive search. BnB complexity is, however, much reduced as it discards parts of the search space wherein the global solution is provably absent.}

Multiple variations of the BnB algorithm exist depending on the method of splitting the search space. However, all these methods rely on the same principle. Given a method to compute an upper and a lower bound on the optimal cost, the algorithm skips a branch of the search space if its lower bound is greater than the upper bound of the other branch. In fact, the optimal solution cannot lie in a branch that satisfies that inequality. From the description above, one can conclude that the performance of the BnB algorithm heavily depends on the accuracy of the upper and lower bound. For example, the lower bound can be obtained by solving a convex relaxation of the problem and the upper bound by using a heuristic. For simplicity purposes, this manuscript focuses on binary optimization and splits the search space canonically by considering the possible values of each binary variable at each branch.

Let $f:\{0,1\}^n \rightarrow \mathds{R}$ be the cost function which maps each vector $\mathbf{x} = (x_1, \ \cdots, \ x_n)$ into a real number. Let $\overline{f}: \mathcal{L} \subseteq \{0,1\}^n \rightarrow \mathds{R}$ and $\underline{f}: \mathcal{L} \subseteq \{0,1\}^n \rightarrow \mathds{R}$ be two functions that return an upper bound and a lower bound, respectively, for any subset $\mathcal{L}$ of the search space. For example, $\overline{f}(0,\ 1,\ x_3, \ \cdots, \ x_n)$ returns an upper bound of the function over all binary vectors whose first and second entries are set to $0$ and $1$, respectively. The BnB algorithm cycles through all variables. For each value of $x_i$ and for each branch $\ell$ in the set of branches $\mathcal{L}$, two new branches are generated for $x_i=0$, denoted by $\ell_0$, and $x_i=1$, denoted by $\ell_1$. The branches $\ell_0$ and $\ell_1$ are added to $\mathcal{L}$ if the solution potentially lies in these branches. In other words, branch $\ell_0$ is added if $\overline{f}(\ell_0) \geq \underline{f}(\ell_1)$, and inversely for $\ell_1$. \fref{fig:branch} shows an example of BnB algorithm applied to a problem of dimension $n=3$ variables. The right sub-tree is ignored as its lower bound is greater than the upper bound of the left sub-tree. The same holds for the left sub-tree generated by the left node. The steps of the algorithm are summarized in \algref{alg3}.

\begin{algorithm}[t!]
\begin{algorithmic}[1]
\REQUIRE Functions $f$, $\overline{f}$, and $\underline{f}$.
\STATE Initialize $\mathcal{L}=\varnothing$.
\FOR{$i=1:n$}
\FORALL{$\ell \in \mathcal{L}$}
\STATE Set $\ell_0 = \ell \cup \{x_i=0\}$
\STATE Set $\ell_1 = \ell \cup \{x_i=1\}$
\IF{$\overline{f}(\ell_0) \geq \underline{f}(\ell_1)$}
\STATE Set $\mathcal{L}=\mathcal{L} \cup \ell_0$
\ENDIF
\IF{$\overline{f}(\ell_1) \geq \underline{f}(\ell_0)$}
\STATE Set $\mathcal{L}=\mathcal{L} \cup \ell_1$
\ENDIF
\ENDFOR
\ENDFOR
\RETURN $\arg\min\limits_{\ell \in \mathcal{L}}f(l)$.
\end{algorithmic}
\caption{Branch-and-Bound Algorithm}
\label{alg3}
\end{algorithm}

\textcolor{black}{While the presented branch-and-bound algorithm is a relatively handy technique for solving the clique problem, its accuracy is strongly coupled with the method chosen to compute the upper and lower bound on the optimal cost. Furthermore, in the worst-case scenario, the complexity of the branch-and-bound algorithm is exponential in the size of the problem. By exploiting the BnB routine, Carraghan and Pardalos \cite{carraghan1990exact} introduced a simple-to-implement algorithm that avoids enumerating all cliques and instead works with a significantly reduced partial enumeration. The algorithm is further improved by \"{O}stergard \cite{16513519} by employing several novel pruning strategies.}

\subsection{\textcolor{black}{Meta-heuristics and Evolutionary Algorithms}}

\textcolor{black}{The heuristics presented in \sref{sec:gra} for solving the clique problem and its variants are universal and can be applied to all graphs. However, they do not offer the opportunity to integrate further information and insights one may have about the problem, e.g., a close to optimal initialization. This part presents additional approaches that exhibit such a feature, namely the meta-heuristics and evolutionary algorithms. Unlike the heuristics in \sref{sec:subqwe} which are specifically designed to solve the clique problem, meta-heuristics, and evolutionary algorithms are higher-level procedures that are independent on the problem under investigation.}

\subsubsection{\textcolor{black}{Meta-heuristics for the Clique Problem}}

\textcolor{black}{As stated earlier, meta-heuristics are high-level procedures to solve a large and diverse collection of problems. Compared to previously mentioned optimization algorithms and iterative methods, meta-heuristics perform stochastic optimization and do not guarantee that a globally optimal solution can be found. However, by searching over a large set of feasible solutions, metaheuristics can often find good solutions with less computational effort than generic methods \cite{bianchi2009survey,blum2003metaheuristics}.}

\textcolor{black}{One particularly successful meta-heuristic for solving the clique problem is the Tabu search (TS) \cite{glover1998tabu}. The main feature of TS is to restrict the feasible space by excluding neighbors that have been visited in the last steps. While local search methods have the tendency to be stuck in local minima, TS outperforms these techniques by prohibiting already visited regions or through user-provided rules. The search history is stored in the form of a tabu list in the form of particular changes of the solution attributes, called moves, that cannot be applied in the subsequent iterations. It is worth mentioning the existence of additional efficient meta-heuristics such as Variable Neighborhood Search (VNS) \cite{hansen2001variable,hansen1999introduction,hansen2010variable}, Variable Neighborhood Descent (VND) \cite{ribeiro2005grasp}, Greedy Randomized Adaptive Search Procedure (GRASP) \cite{feo1989probabilistic,feo1995greedy}, and a combination of these methods \cite{salehipour2011efficient}. Interested readers are directed to the following references for more information about meta-heuristics \cite{glover2006handbook}, their implementations \cite{talbi2009metaheuristics}, and distributed solutions \cite{alba2005parallel}.}

\subsubsection{\textcolor{black}{Evolutionary Algorithms}}

Stochastic algorithms are widely used in artificial intelligence, thanks to their stochasticity, which helps deal with various graph structures. A survey on the available stochastic algorithms can be found in \cite{maffioli1986randomized,srinivas1994genetic,cantu1998survey}. This section is interested in a particularly popular class of algorithms known as Evolutionary Algorithms (EAs) \cite{vikhar2016evolutionary}.

EAs are generic population-based optimization algorithms inspired by biological evolution, such as reproduction, mutation, recombination, and selection. Generally, feasible solutions to the optimization problem play the role of individuals in a population that evolves to achieve better solutions. The evolution of the population depends on the used EA. For example, the bee algorithm  \cite{pham2006bees}, first developed by Pham et al. \cite{pham2005bees}, mimics the food foraging behavior of honey bee colonies by performing a neighborhood search combined with a global search. The effectiveness of the bees algorithm is proven in some applications \cite{nasrinpour2017grouped,pham2014benchmarking,pham2015comparative}.

\textcolor{black}{EA algorithms have been employed to solve the clique problem in various settings. For example, the authors in \cite{6557756} design a memetic algorithm for the maximum clique problem, i.e., an EA augmented with a local search. Similarly, Binary Particle Swarm Optimization (BPSO) is an important instance of EA with a robust implementation, e.g., Leapfrogging \cite{manimegalai2014improved,rhinehart2014convergence,rhinehart2012leapfrogging
}, and fixed computation complexity. Indeed, optimal algorithms are of less importance in situations where the computation power is limited, and/or where a result needs to be returned after a given number of iterations. Eberhart and Kennedy introduce the Particle Swarm Optimization (PSO) as a search algorithm \cite{65182525,637339} that simulates the social behavior of animals, e.g., birds. The fundamental concept of BPSO \cite{4433821,44335621} is to generate $L$ individuals in the $N$ multi-dimensional space, called herein particles. Each particle evolves according to its position and its velocity. The best position all the particles visit after $T$ iterations is referred to as the all-best position and is returned by the algorithm. A comparison of the performance of the different EAs can be found in \cite{rodriguez2016performance}.}

\section{Applications of the Maximum Clique Problem in Communications and Signal Processing} \label{sec:gra2}

\textcolor{black}{The previous part of this manuscript explains the variants of the clique problem, and sheds lights on their optimal and heuristic solutions. The rest of the paper illustrates the clique formulation applications in communications and signal processing problems. We present few applications of the maxim clique problem in this section. The applications of the maximum weight clique problem and the $k$-clique problem are discussed in the next subsequent sections, respectively.}

The maximum clique problem is one of the most studied problems in graph theory thanks to its numerous applications notably in communications and signal processing, e.g., circuit design \cite{35554}, algebraic code design \cite{7423981}, quantum error correcting codes \cite{6544815}, network coding \cite{14040265,douik2018data,7565067}, parallel computing \cite{8425226}, and constraint satisfaction problems \cite{5171273,qin2009maximum,vismara2008finding}.

Programmable Logic Arrays (PLAs) are widely used structured logic arrays that implement multiple output combinational logic functions. In general, a PLA can be described in a symbolic form by a sparse matrix called a personality matrix. A direct implementation of a PLA results in a significant waste of chip area due to the sparsity of the personality matrix, which results in a reduction of the circuit yield and degradation of the PLA's time performance. PLA folding \cite{hachtel1982algorithm,wong1988pla} is a general technique to reduce the total area of a PLA by letting two input variables or two output functions sharing the same column. The authors in \cite{35554} exploit the clique formulation to solve the PLA folding problem. In particular, the authors transform the PLA into graphs wherein cliques correspond to PLA folding sets. A simple heuristic algorithm is proposed to solve the problem, and through experimental testing, the algorithm is demonstrated to identify near-maximum cliques in polynomial time.

\textcolor{black}{The clique formulation has been successfully used to solve algebraic coding problems \cite{montemanni2015combinatorial}. For example, permutation codes, also known as permutation arrays, gained a lot of attention in recent years thanks to their real-world applications \cite{dukes2010bounds,bogaerts2010new,gao2013improvement}, e.g., in powerline communications \cite{pavlidou2003power,colbourn2004permutation}. Permutation codes ensure that power output remains as constant as possible in the presence of noise. The authors in \cite{7423981} address the problem of constructing the largest possible permutation codes with a specified length and minimum Hamming distance. The problem is reformulated as a maximum clique search in a well-designed graph, and a BnB method is tailored for the design of permutation codes.}

Parallel computing networks can be represented by a graph $\mathcal{G}(\mathcal{V},\mathcal{E})$ wherein each vertex $v \in \mathcal{V}$ in the graph is a processor and each edge $\epsilon_{ij} \in \mathcal{E}$ is a communication link between the processors $v_i$ and $v_j$. The authors in \cite{8425226} employ the clique formulation to derive a distributed symmetry breaking \cite{fraigniaud2016local} in graphs with bounded diversity wherein the diversity of a graph is defined as the maximum number of maximal cliques a vertex belongs to. The authors in \cite{fraigniaud2016local} further characterize the running time of their algorithm for various common graph structures.

Constraint satisfaction problems are problems in which one desires to attest to the existence of an object that satisfies a number of constraints. These problems are also known as the satisfiability problem with the Boolean satisfiability problem (SAT) being the most famous as it is the first problem to be shown to be NP-complete. Graph theory and the SAT played a major role in demonstrating the intractability of multiple problems. In particular, the clique formulation has been employed in \cite{5171273} and \cite{qin2009maximum} to show equivalency between clause sets and undirected graphs in polynomial time, such that finding a renamable Horn subset of a clause set is equivalent to finding an independent set of vertices of a graph. Similarly, Vismara and Valery \cite{vismara2008finding} propose finding the maximum common connected subgraphs using clique detection and constraint satisfaction algorithms.

The rest of this section investigates in detail the use of the maximum clique formulation to solve two problems in communications and signal processing. The first part of the section illustrates the use of the clique formulation to enhance the channel use, i.e., maximum access, in a non-orthogonal multiple access (NOMA) network. The second subsection formulates the problem of reducing the number of transmissions using index coding, a network coding scheme, as a clique search over a well-defined graph.

\subsection{Maximum Access Problem for Non-Orthogonal Multiple Access Networks}

The urgent demand for increased spectrum efficiency and network capacity of the next generation wireless systems (5G and beyond) \cite{6824752} requires the optimization of traditional communication techniques. Non-orthogonal multiple access (NOMA) networks are proposed as a potential solution thanks to their superior capabilities as compared to traditional orthogonal multiple access schemes \cite{dai2015non,tse2005fundamentals}. Indeed, unlike traditional orthogonal communication techniques, NOMA successfully serves multiple users using the same spectrum by exploiting the user diversity in the power domain and decoding the overlapping signals using the successive interference cancellation technique.

Owing to its aforementioned benefits, resource management in NOMA networks has recently received considerable attention in the research community \cite{lei2016power,choi2016power,zhai2018energy,zhang2016uplink,zhai2018spectrum,
ruby2018enhanced}. For example, the authors in \cite{lei2016power,choi2016power} focus on optimizing the resource allocation in the downlink of a NOMA network by designing power allocation schemes for SISO-NOMA and MIMO-NOMA networks, respectively. Reference \cite{zhai2018energy} includes dynamic user scheduling with power allocation in their algorithm. Similarly, the authors in \cite{zhang2016uplink,ruby2018enhanced,zhai2018spectrum} focus on improving the data rate and power consumption in the uplink transmissions of a NOMA network.

All aforementioned works on resource management and optimization in NOMA are conducted under the assumption that the network resources are sufficient enough to accommodate all users. Such an assumption, however, does not hold in the Internet of Things (IoT)-based systems, which aims at connecting a massive number of devices to a limited amount of radio resources. In such systems, NOMA techniques are not able to support all of the devices simultaneously. Therefore, an admission control policy needs to be designed concurrently with resource allocation algorithms.

This part studies the maximum access problem for non-orthogonal multiple access networks. In particular, the resource allocation problem is formulated as an integer program with admission control, user clustering, and channel assignment. The next part of the tutorial introduces the considered system model and formulates the resource allocation problem in non-orthogonal multiple access networks. Afterward, a graph is designed in such a way that each clique represents a feasible solution. The problem is then formulated as a search for the maximum independent set in the designed graph.

\subsubsection{System Model and Parameters}

Consider the uplink of a non-orthogonal multiple access network consisting of a set $\mathcal{U}$ of $U$ users and a set $\mathcal{K}$ of $K$ channels. Let $h_{uk}$ represent the complex channel gain between the $u$-th user and the $k$-th channel. The channel power gain (CPG), defined as $|h_{uk}|^2$ for the $u$-th user and the $k$-th channel, plays an important role in NOMA networks. Indeed, the optimal decoding order in uplink NOMA transmissions is the decreasing order of the CPGs. In other words, interference for a particular user is seen from other users who both share the same channel and have a lower CPG. Decoding is achieved using successive interference cancellation (SIC). Due to the complexity of SIC in practical systems, this tutorial assumes that each channel can accommodate at most two users, e.g., see \cite{zhai2018spectrum,8365765}.

Let $P_u$ be the transmit power of the $u$-th user, which is assumed to be fixed to a nominal value. Define the binary variable $a_u$ with $a_u = 1$ is the $u$-th user is allowed to access the channel and $0$ otherwise. Similarly, let $s_{uk}$ be a variable that indicates if the $u$-th user is scheduled to the $k$-th channel. The signal-to-interference plus noise-ratio (SINR) experienced by the $u$-th user when connected to the $k$-th channel can then be expressed as follows:
\begin{align}
\gamma_{uk} = \cfrac{P_{u} a_u s_{uk}|h_{uk}|^2}{\Gamma\left(\sigma^2+ \sum\limits_{\{u^\prime \ | \ |h_{u^\prime k}|^2 \leq |h_{uk}|^2\}}P_{u^\prime} a_{u^\prime} s_{u^\prime k} |h_{u^\prime k}|^2\right)},
\end{align}
where $\sigma^2$ is the Gaussian thermal noise variance, and $\Gamma$ denotes the SINR gap from Shannon capacity which accounts for the use of finite length codewords and practical constellations.

Let $\text{R}_{uk}$ denote the data-rate of the $u$-th user when connected to the $k$-th channel, and let $\text{R}_{u}$ be the total achievable capacity for the $u$-th user. The data rate can be expressed as
\begin{align}
\text{R}_{u} = \sum_{k \in \mathcal{K}} \text{R}_{uk} = \sum_{k \in \mathcal{K}} \mathcal{B}_{uk} \log_2(1+\gamma_{uk}),
\end{align}
wherein $\mathcal{B}_{uk}$ denotes the available bandwidth.

\subsubsection{Joint Admission Control and Resource Allocation}

Under the above described NOMA network, the base-station is responsible for scheduling users to each channel under the constraint that each user can connect to exactly single channel and that each channel can accommodate at most a couple of users. Furthermore, to guarantee a certain quality of service, each user allowed to access the channel, i.e., $a_u = 1$, should be guaranteed a certain minimum rate $\text{R}_{u}^{\min}$. Therefore, the aim is to maximize the number of served users by allocating the channels under strict rate requirement as shown in the following system constraints:
\begin{itemize}
\item \textbf{C}1: The rate of the $u$-th user should exceed the nominal value $a_u \text{R}_{u}^{\min}$.
\item \textbf{C}2: Each user can connect to exactly a single channel.
\item \textbf{C}3: Each channel can serve a maximum of $2$ users.
\end{itemize}
Given the definition of the binary variables $a_u$ and $s_{uk}$, the joint admission control and resource allocation problem in NOMA networks can be formulated as follows
\begin{subequations}
\label{Original_optimization_problem5874}
\begin{align}
\max_{a_u, s_{uk} \in \{0,1\}} & \sum_{u \in \mathcal{U}} a_u \\
\label{Originalx1} {\rm s.t.\ }  &\text{R}_{u} \geq a_u \text{R}_{u}^{\min} , \forall \ u \in\mathcal{U},\\
\label{Originalx2} & \sum_{k \in \mathcal{K}} a_u s_{uk} = 1,\quad \forall\ u\in\mathcal{U},\\
\label{Originalx3} & \sum_{u \in \mathcal{U}} a_u s_{uk} \leq 2, \quad\forall\ k\in\mathcal{K},
\end{align}
\end{subequations}
where the optimization is over the binary variables $a_u$ and $s_{uk}$ for all $(u,k) \in \mathcal{U} \times \mathcal{K}$. In the optimization problem \eref{Original_optimization_problem5874}, constraint \eref{Originalx1} translates the system constraint \textbf{C}1. Likewise, constraints \eref{Originalx2} and \eref{Originalx3} correspond to system constraint \textbf{C}2 and \textbf{C}3.

While the power levels $P_u$ are considered fixed in the formulation above, the analysis can be extended to optimize the power allocation with an extra iterative step for computing the optimal power distribution, e.g., see \cite{zhai2018spectrum,8365765} for a joint resource and power allocation for multi-carrier in non-orthogonal multiple access networks. Even though the resulting problem in \cite{zhai2018spectrum,8365765} is a mixed continuous-integer program, it can be solved using the maximum clique formulation. This fact is further explored in \sref{sec:gra5}, albeit in a different context.

\subsubsection{Maximum Independent Set Reformulation}

This part designs a graph $\mathcal{G}(\mathcal{V},\mathcal{E})$ such that each feasible solution to \eref{Original_optimization_problem5874} corresponds to an independent set in that graph. The tutorial first designs the set of vertices $\mathcal{V}$ such that all vertices satisfy the system constraint \textbf{C}1. Afterward, the set of edges $\mathcal{E}$ is constructed such that the system constraints \textbf{C}2 and \textbf{C}3 hold. The below graph formulation for the resource allocation problem in NOMA network is first proposed in \cite{zhai2018spectrum} for joint spectrum-efficiency resource management and power allocation.

Let $\mathcal{Q}$ denote a user cluster wherein all users share the same channel. From the system constraint \textbf{C}3, the set $\mathcal{Q}$ contains at most two users, i.e., $\mathcal{Q}$ is an element of the set $\overline{\mathcal{U}} = \mathcal{U} \times \{\mathcal{U} \cup \varnothing \}$. Given a channel $k$ and a cluster $Q$, the assignment is feasible only if the rate constraint \textbf{C}1 is satisfied, i.e., for all $u \in \mathcal{Q}$, we have $\text{R}_{u} \stackrel{(a)}{=} \text{R}_{uk} \geq \text{R}_{u}^{\min}$ with the equality $(a)$ obtained from the system constraint \textbf{C}2. The set of all feasible users and channel assignment is given by the set $\mathcal{A}$ defined as
\begin{align}
\mathcal{A} = \{(\mathcal{Q},k) \in \overline{\mathcal{U}} \times \mathcal{K}\ | \text{R}_{uk} \geq \text{R}_{u}^{\min},\ \forall \ u \in \mathcal{Q} \}
\end{align}

Let $\mathcal{G}(\mathcal{V},\mathcal{E})$ be an undirected graph representing all allocations of users and channels. The set of vertices is generated by creating a vertex $v \in \mathcal{V}$ for all possible assignments $a \in \mathcal{A}$. To design the set of edges $\mathcal{E}$, the tutorial introduces the mapping functions $\varphi_u$ and $\varphi_k$ from the set $\mathcal{A}$ to the set $\overline{\mathcal{U}}$ and $\mathcal{K}$, respectively, such that for any $a = (\mathcal{Q},k) \in \mathcal{A}$, we have $\varphi_u(a)=\mathcal{Q}$ and $\varphi_k(a)=k$. Two vertices $v$ and $v^\prime$ are connected if one of the following conditions holds
\begin{itemize}
\item $\varphi_u(v) \cap \varphi_u(v^\prime) \neq \varnothing$: This condition denotes that the same user is scheduled to different channels, which violates the system constraint \textbf{C}2.
\item $\varphi_k(v) = \varphi_k(v^\prime)$: This condition indicates that the same channel is allocated to more than $2$ users, which violates \textbf{C}3.
\end{itemize}

By the construction of the set $\mathcal{A}$, it is clear that the system constraint \textbf{C}1 is satisfied for all vertices. Furthermore, it is not difficult to see that the independent set corresponds to feasible solutions to the problem \eref{Original_optimization_problem5874} as all vertices in the independent set satisfy conditions \textbf{C}2 and \textbf{C}3. Finally, given that the objective function of \eref{Original_optimization_problem5874} corresponds to the size of the independent set in the graph $\mathcal{G}(\mathcal{V},\mathcal{E})$ as designed above, one can conclude that the maximum access problem in NOMA is equivalent to a maximum independent set search.

\subsection{Throughput Maximization Using Index Coding}

We now illustrate another example that can be solved using the maximum clique problem. In their seminal paper \cite{850663}, Ahlswede, Cai, and Yeung suggest mixing different information flows within intermediate nodes in the network, thereby increasing the information content per transmission \cite{19518321}. Using the butterfly network as an example, the authors proved that Network Coding (NC) could enhance the transmission efficiency and throughput up to the network capacity. With such promising capabilities, network coding has been adopted by many researchers worldwide, e.g., \cite{5669233,5506109}, to design high data rate communication algorithms. The schemes are further developed to meet the constraints of different communication scenarios, e.g., improve security, quality of service, manageability, and robustness.

Thanks to their combinatorial nature, problems in NC can often be formulated as graph theory problems in well-designed graphs. One of the most famous examples of such formulation is the max-flow-min-cut theorem \cite{19518321}, which allows deriving the maximum flow passing from a source to a sink as the minimum cut in the connectivity graph, where the capacity of each link is represented as the weight of the associated edge.

This tutorial focuses on a particular subclass of network coding known as the index coding (IC) and introduced by Birk and Kol \cite{1638566}. Unlike general NC schemes, index coding provides instantaneous and low-complexity encoding and decoding strategies, which are suitable for battery-powered devices. Such a goal is achieved by solely using XOR operations to encode and decode files. These simple properties lead to substantial progress in the design and analysis of index coding schemes in various network settings \cite{4313060,4595303,4544612,10041379,6283912,2095151,6620405,6283853,6283851}, including index coding in directed graph \cite{7746676}.

\subsubsection{System Description and Problem Formulation}

Consider a set $\mathcal{U}$ of $U$ users interested in receiving a set $\mathcal{F}$ of $F$ files from the transmitting base station. Each user possesses some part of $\mathcal{F}$, known as its side information, that it receives in previous transmissions. The side information of the $u$-th user is modeled by the following couple of sets:
\begin{itemize}
\item The ``\textit{Has}" set $\mathcal{H}_u$ represents the files available at the $u$-th device.
\item The ``\textit{Wants}" set $\mathcal{W}_u$ represents the files missing at the $u$-th device.
\end{itemize}

\begin{figure}[t]
\centering
\includegraphics[width=\linewidth]{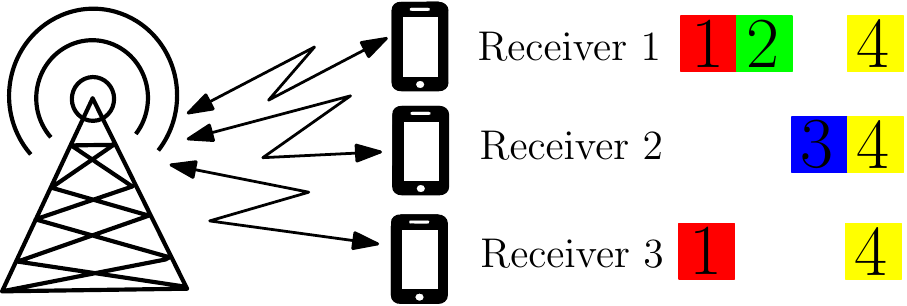}\\
\caption{The base-station is required to deliver $4$ packets to $3$ users. User $1$ is missing packet $3$, user $2$ is missing packets $1$ and $2$ and user $3$ is missing packets $2$ and $3$.}\label{fig:8}
\end{figure}

From the definition above, the \textit{Has} and \textit{Wants} sets are clearly complementary, i.e., $\mathcal{F} = \mathcal{H}_u \cup \mathcal{W}_u$, $\forall$ $u \in \mathcal{U}$. \fref{fig:8} shows an example that illustrates the benefit of IC through a base-station that aims at delivering four files to three users. If only uncoded transmissions are allowed, the base-station would require three-time slots to serve all devices. Using IC, however, only $2$ transmissions would be required since, based on the distribution of the \textit{Wants} set, the transmission of file combination $1 \oplus 3$ would be beneficial to all three users. In the second time slot, the uncoded second file is re-transmitted.

The index problem of interest herein is to transmit the file combination that is decodable by all users, and that contains a maximum number of source files. Let $\kappa \in \mathcal{P}(\mathcal{F})$ be the file combination to be transmitted, where the notation $\mathcal{P}(\mathcal{X})$ refers to the power set of the set $\mathcal{X}$, i.e., the set of all subsets of $\mathcal{X}$. A packet $\kappa$ is said to be decodable by the $u$-th user if it contains at most a single missing packet, i.e., $|\kappa \cap \mathcal{W}_u| \leq 1$. Similarly, for a file combination $\kappa$, the number of source packets it contains is $|\kappa|$. Note that the transmission $\kappa$ is beneficial to the $u$-th user if it contains exactly a single missing packet, i.e., $\ |\kappa \cap \mathcal{W}_u| = 1$. Indeed, due to the exclusive use of binary XOR, only users for which the transmission $\kappa$ is beneficial can recover one of their missing files by XORing the received package with some files in their Has set.

\textcolor{black}{Therefore, the IC coding problem can be formulated as an optimization problem wherein one determines the appropriate file combinations so as to maximize the file diversity of the packet. In other words, the index coding problem can be formulated as
\begin{subequations}
\label{eq:87r1}
\begin{align}
&\max_{\kappa \in \mathcal{P}(\mathcal{F})} |\kappa| \\
&\text{s.t. }  |\kappa \cap \mathcal{W}_u| \leq 1, \ \forall \ u \in \mathcal{U},
\end{align}
\end{subequations}
where the maximization is over the potential file combinations $\kappa$, and the constraint insists that the combination is decodable by all users.}

The solution to the above problem would usually require an exhaustive search over all possible file combinations so as to determine the combination that provides the highest objective function. The complexity of such a process would scale as $2^F$, which is clearly non-feasible for any reasonably sized set of files. The rest of the section uses the celebrated clique formulation to solve the above problem with reasonable computational complexity.

\subsubsection{Optimal Index Coding Packet Combination}

The index coding graph $\mathcal{G}(\mathcal{V},\mathcal{E})$ is a graphical representation of all possible message combinations. From the problem formulation in \eref{eq:87r1}, a candidate for the set of vertices is the set $\mathcal{F}$. In other words, the IC graph $\mathcal{G}(\mathcal{V},\mathcal{E})$ is constructed by generating a vertex $v_{f} \in \mathcal{V}$ for file $f \in \mathcal{F}$. Therefore, to formulate the problem as a maximum clique search, one needs to design a suitable set of connections.

As stated in \sref{sec:opt}, one needs to find a set of edges $\mathcal{E}$ such that each feasible solution for \eref{eq:87r1} corresponds to a clique in the graph $\mathcal{G}(\mathcal{V},\mathcal{E})$. Intuitively, two packets, i.e., vertices, can be combined if and only if they are decodable at all users. To express the condition mathematically, we first define the set $\mathcal{U}_f$ for all files $f \in \mathcal{F}$ such that it represents all users that are interested in the file $f$, i.e., $\mathcal{U}_f = \{u \in \mathcal{U} \ | \ f \in \mathcal{W}_u\}$. Given the above definition, two vertices $v_f$ and $v_{f^\prime}$ are connected by an edge if and only if $\mathcal{U}_f \cap \mathcal{U}_{f^\prime} = \varnothing$.

\begin{figure}[t]
\centering
\includegraphics[width=0.4\linewidth]{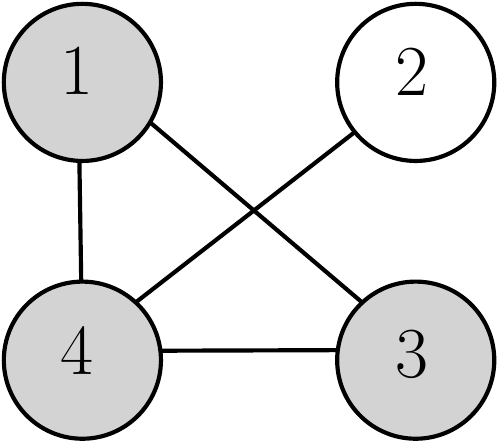}\\
\caption{The IC graph representing the coding possibilities for the system depicted in \fref{fig:8}. Each clique in the graph represents a feasible packet combination. In particular, one can see that vertices $1$, $3$, and $4$ represent a clique corresponding to transmitting the combination $1 \oplus 3 \oplus 4$.}\label{fig:ss10}
\end{figure}

Given the above method of edge generation, it is clear that each clique in the IC graph corresponds to a feasible message combination. Similarly, each feasible combination can be represented by a clique in the IC graph. Finally, as the objective function of \eref{eq:87r1} corresponds naturally to the size of the clique in the IC graph, it is clear that solving \eref{eq:87r1} is equivalent to finding the maximum clique in the graph $\mathcal{G}(\mathcal{V},\mathcal{E})$. For illustration purposes, the IC graph of the system available in \fref{fig:8} can be found in \fref{fig:ss10}. Recasting problem \eref{eq:87r1} as a clique search allows us to solve the problem using state of the art techniques efficiently.

While the complexity of solving the maximum weight clique is intrinsic to the problem, the complexity of constructing the IC graph is not. In fact, one can take advantage of integer optimization methods to solve the problem either optimally or efficiently. Multiple specialized stochastic algorithms, i.e., heuristics, are further available in the literature, e.g., the base policy-based partial enumeration (BPPE) algorithm proposed in \cite{8002629} for packet finding.

\section{Applications of the Maximum Weight Clique Problem in Communications and Signal Processing} \label{sec:gra3}

\textcolor{black}{The maximum weight clique problem is an extension of the maximum clique wherein nodes are allowed to have weights. Therefore, similar to the maximum clique problem, the maximum weight clique problem has numerous applications in communications and signal processing, e.g., see references \cite{6848102} and \cite{pattillo2011clique} for applications of the weight clique problem in communications.}

The maximum weight clique problem appears naturally in scheduling problems wherein nodes have different payoffs for being selected. For example, the authors in \cite{8126874} suggest an efficient spectrum resource management for multi-carrier non-orthogonal multiple access networks. Specifically, the authors consider the decoding threshold of successive interference cancellation and formulate the problem as a sum-rate maximization problem by jointly considering the user pairing, channel assignment, and power control. After manually setting the optimal power-levels, the authors show that the problem is equivalent to a maximum weight clique problem that can be efficiently solved using their proposed heuristic. Likewise, reference \cite{8450715} uses the clique formulation to design a distributed transmission scheduling algorithm for wireless networks using the Ising model.

Other applications of the maximum weight clique problem in communications is the design of 5G enabled vehicular ad hoc network. In such networks, many applications rely on efficient content sharing among mobile vehicles, which is a challenging issue due to the extremely large data volume, rapid topology change, and unbalanced traffic. The authors in \cite{8424578} investigate content pre-fetching and distribution in VANETs, which can improve the sharing efficiency and alleviate the burden on the network. The content distribution problem is formulated as a maximum weighted independent set search, and its performance and efficiency are attested through extensive simulations.

Besides the aforementioned applications of the maximum weight clique in communications, the problem has been successfully implemented to solve numerous signal processing applications such as video technology. For example, the authors in \cite{8533372} propose a novel iterative maximum weighted independent set algorithm for multiple hypotheses tracking in a tracking-by-detection framework. The multiple hypothesis tracking problem is converted into a series of clique problems across the tracking time. While previous attempts solve each clique problem separately, the authors in \cite{8533372} suggest using the solution from the previous frame to solve the next rather than solving the problem from scratch each time, which allows the solution to outperform all previously published tracking algorithms. On the other hand, reference \cite{7088570} aims to design an efficient action detection method for real-world videos. Given a space-time graph representing the entire action video, the proposed method identifies a maximum weight clique indicating the detection of an action. From extensive experimentation with real-world data sets, it is concluded that the clique-based solution provides more accurate localization as compared with conventional methods.

The rest of the section mathematically illustrates the use of the maximum weight clique formulation to solve a couple of problems in communications and signal processing. The first part of the section extends the index coding problem to a more general setting known as instantly decodable network coding \cite{7414149,douik2018delay,164579}. The decoding delay problem is then formulated and shown to be equivalent to a maximum weight clique search in the appropriate graph. The paper afterward highlights another application that focuses on designing collision-free radio frequency identification networks.

\subsection{Instantly Decodable Network Coding}

Instantly decodable network coding (IDNC) is a particular subclass of network coding that generalizes index coding. While links are assumed to be perfect in index coding, IDNC relaxes the assumption by considering erasure prone links. However, similar to IC,  IDNC provides instantaneous and low-complexity encoding and decoding strategies through the exclusive use of binary XOR operations to encode and decode files, which are suitable for battery-powered devices. These simple properties lead to substantial progress in the design and analysis of IDNC schemes in various network settings \cite{14122014,13051412,445913,6673389,665497}. A survey on the recent advances and applications of IDNC is available in \cite{7845689}.

This section illustrates how a maximum weight clique search is exploited to solve the user maximization IDNC problem using a graph model to represent all encoding strategies at the sender. The first part introduces the system parameters and formulates the problem. The second part introduces the encoding graph, known as the IDNC graph, and shows that the problem is equivalent to a maximum weight clique search in that graph.

\subsubsection{Problem Statement and Formulation}

\textcolor{black}{Consider a set $\mathcal{U}$ of $U$ users interested in receiving a set $\mathcal{F}$ of $F$ files from the transmitting base station. In an initial phase, the wireless base-station broadcasts each file of the set $\mathcal{F}$ sequentially. Due to the dynamic nature of the channel, e.g., fading, shadowing, and thermal noise, some of these files are erased (lost) at some users. Users that successfully receive a file acknowledge its reception by sending an ACK. Thanks to their small size, the transmission of these ACKs is assumed to be erasure-free, which can be achieved using dedicated feedback channels and employing error-correcting coding schemes. Erasures are modeled as independent and identically distributed Bernoulli random variables. Let $\epsilon_u$ be the erasure probability of the transmission from the base station to the $u$-th user. The value of $\epsilon_u$ is assumed to be constant during the transmission of a single file $u\in \mathcal{U}$. Furthermore, the tutorial adopts the standard assumption that the $\epsilon_u$'s are perfectly available at the transmitter.}

\textcolor{black}{The decoding delay is defined in the following fashion. Each user $u$ with non-empty \textit{Wants} set experiences one unit of decoding delay $\mathcal{D}_u =1$ if it successfully receives a message that does not allow it to reduce its Wants set.  Therefore, if the transmission at the $u$-th user is erased, no decoding delay is incurred. However, if the transmission is successful, the user receives one unit of decoding delay if he is not targeted by the transmission. Therefore, the expected delay is $\mathbb{E}(\mathcal{D}_u) = (1-\epsilon_u)$ if the user is not targeted and $0$ otherwise. Let $\overline{\mathcal{U}} \subseteq \mathcal{U}$ be the set of users with non-empty Wants set. At each transmission slot, the base-station aims to transmit the file combination that has the minimum expected decoding delay \cite{5683677}, i.e., the combination which minimize $\mathbb{E}(\sum_{u \in \overline{\mathcal{U}}} \mathcal{D}_u)$.}

\textcolor{black}{Let $\kappa \in \mathcal{P}(\mathcal{F})$ be the file combination to be transmitted. From the IDNC limitations, the set of targeted users by the file combination $\kappa$, defined as $\tau(\kappa)$, should contain a single missing file, i.e., $\ |\kappa \cap \mathcal{W}_u| = 1$, $\forall \ u\in \tau(\kappa)$. Let $\overline{\tau}(\kappa) = \overline{\mathcal{U}} \setminus \tau(\kappa)$ be the set of non-targeted users with non-empty Wants set. The optimization problem considered in this tutorial consists of choosing the appropriate file combination to result in the smallest collective decoding delay. Such optimization problem can be formulated as:
\begin{subequations}
\begin{align}
&\min_{\kappa \in \mathcal{P}(\mathcal{F})} \sum_{u \in \overline{\tau}(\kappa)} (1-\epsilon_u) \\
&\text{s.t. } \tau(\kappa) = \left\{u \in \mathcal{U} \ \Big| \ |\kappa \cap \mathcal{W}_u| = 1\right\} \\
& \hspace{0.6cm} \overline{\tau}(\kappa) = \overline{\mathcal{U}} \setminus \tau(\kappa)
\end{align}
\end{subequations}
which is equivalent to the following formulation (see \cite{5683677} for more details):
\begin{subequations}
\label{eq:1}
\begin{align}
&\max_{\kappa \in \mathcal{P}(\mathcal{F})} \sum_{u \in \tau(\kappa)} (1-\epsilon_u) \\
&\text{s.t. } \tau(\kappa) = \left\{u \in \mathcal{U} \ \Big| \ |\kappa \cap \mathcal{W}_u| = 1\right\}
\end{align}
\end{subequations}
where the maximization is over the potential file combinations $\kappa$. Similar to index coding, the solution to the above problem require an exhaustive search over all possible file combinations so as to determine the combination that provides the highest objective function, which is clearly non-feasible. The rest of the section uses the celebrated IDNC graph formulation \cite{5683677}, and solves the above problem with a reasonable computational complexity.}

\subsubsection{IDNC Graph and Maximum Weight Clique}

\begin{figure}[t]
\centering
\includegraphics[width=0.4\linewidth]{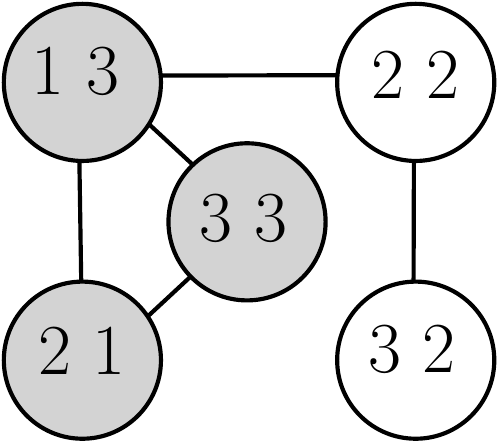}\\
\caption{The IDNC graph representing the coding possibilities for the system depicted in \fref{fig:8}. Each clique in the graph represents a feasible packet combination. In particular, one can see that vertices $13$, $21$, and $33$ represent a clique corresponding to transmitting the combination $1 \oplus 3$.}\label{fig:10}
\end{figure}

\textcolor{black}{The IDNC graph is a tool that represents all coding possibilities and targeted users; see \cite{5683677} and references therein. For example, the IDNC graph of the system illustrated in \fref{fig:8} can be found in \fref{fig:10}. The graph has been extended to fit multiple communication scenarios, e.g., the authors in \cite{6774841} introduce a lossy version of the graph for imperfect feedback systems. Likewise, a multi-layer IDNC graph is proposed in \cite{7500144} to prioritize critical users, and the authors in \cite{7769211} suggest a rate-aware IDNC graph that incorporates the transmission rate in the coding decisions.}

The IDNC graph $\mathcal{G}(\mathcal{V},\mathcal{E})$ is constructed by generating a vertex $v_{uf} \in \mathcal{V}$ for each device $u \in \mathcal{U}$ and each of its missing files $f \in \mathcal{W}_u$. The edges are generated such that connected users can decode simultaneously the file combinations represented by the vertices. From the IDNC constraints, an edge is generated between a couple of vertices $v_{uf}$ and $v_{u^\prime f^\prime}$ if and only if one of the following connectivity conditions (\textbf{CC}) is true:
\begin{itemize}
\item \textbf{CC1}: $f = f^\prime$. This condition implies that the same file is requested by both users. Therefore, both can be served simultaneously.
\item \textbf{CC2}: $f \in \mathcal{H}_{u^\prime}$ and $f^\prime \in \mathcal{H}_{u}$ . This condition states that the missing file at each user is available at the other and vice-versa. Therefore, by XORing the combination with the appropriate file in their \textit{Has} sets, both users $u$ and $u^\prime$ can be targeted at the same time.
\end{itemize}

Based on the file recovery requirements, a feasible file combination is a combination that can help at least a single user recovering one of its missing files. From the graph construction, each clique in the graph represents a file combination that can be decoded by all users represented by the corresponding vertices and, thus, points to a feasible combination. On the other hand, a contrapositive argument would allow showing that any feasible file combination for some users is necessarily represented by a clique in the IDNC graph. Therefore, there is a one-to-one mapping between the clique of the graph and the feasible file combinations. Furthermore, one can see that the set of targeted devices $\tau(\kappa)$ by a combination $\kappa$ represented by a \emph{maximal} clique in the IDNC graph is the set of users identified by the vertices in that clique. These two simple properties of the IDNC graph above allow to rewrite the optimization problem \eref{eq:1} as follows:
\begin{align}
\max_{C \in \mathcal{C}} \sum_{u: v_{uf} \in C} (1-\epsilon_u),
\end{align}
wherein $\mathcal{C}$ is the set of all maximal cliques in the IDNC graph. Furthermore, by defining the weight of each vertex $v_{uf} \in \mathcal{V}$ as $\omega(v_{uf}) = 1-\epsilon_u$, one can rewrite the problem using solely the graph parameters (i.e., cliques, vertices, and weights):
\begin{align}
\max_{C \in \mathcal{C}} \sum_{u: v_{uf} \in C} \omega(v_{uf}).
\end{align}

\begin{figure}[t]
\centering
\includegraphics[width=\linewidth]{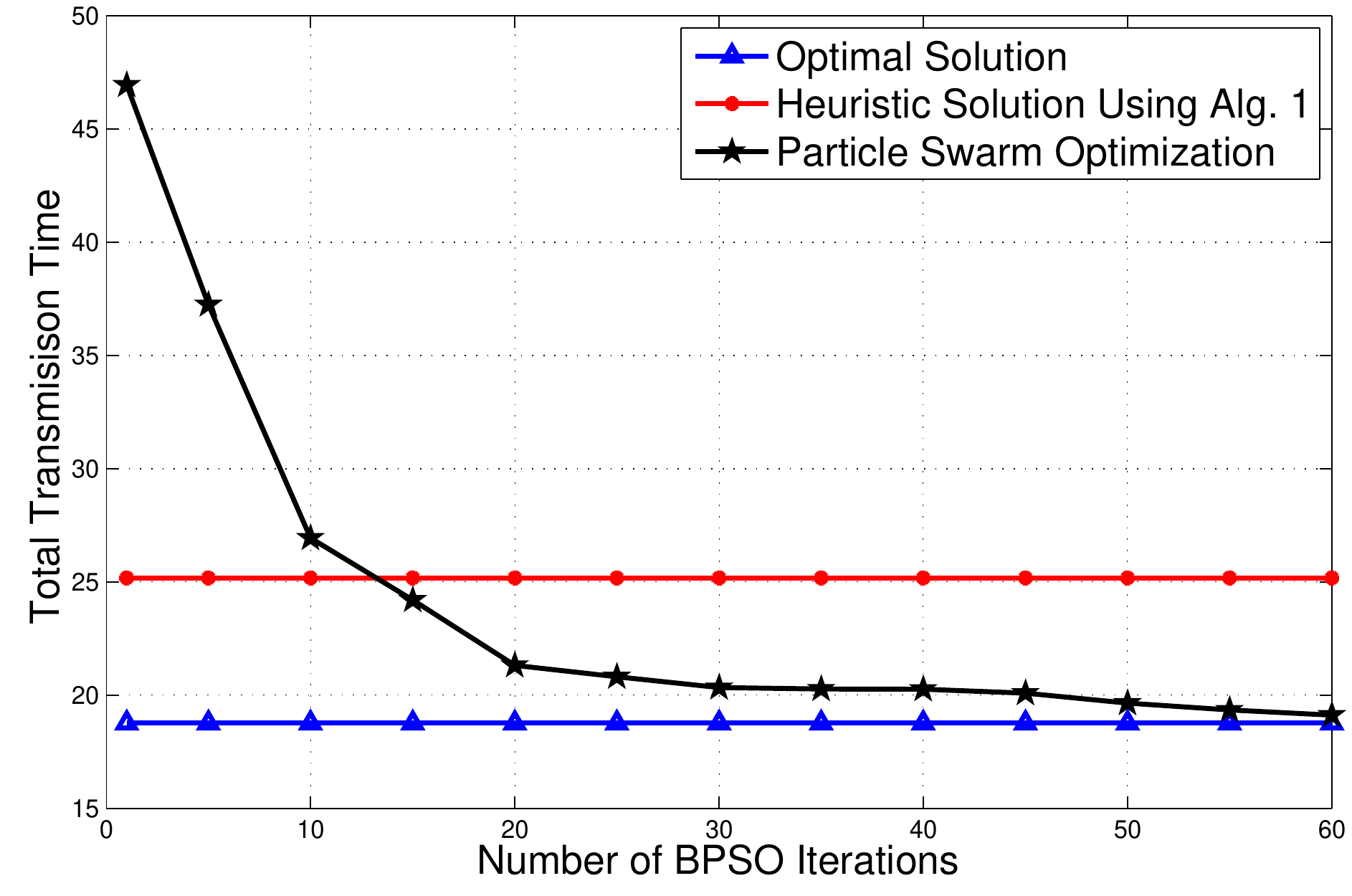}\\
\caption{Total transmission time versus the number of BPSO iterations for a network composed of $30$ users, $30$ packets and an erasure probability $\epsilon =0.1$.}\label{fig:rwy}
\end{figure}

\textcolor{black}{The above reformulation of problem \eref{eq:1} implies that the user maximization IDNC problem is equivalent to a maximum weight clique search in the corresponding IDNC graph. Therefore, optimal and efficient algorithms from literature can be used to find the solution in a much more efficient manner than the exhaustive search method. \fref{fig:rwy} illustrates the total time required to complete the transmission of $30$ files at $30$ users using IDNC in a wireless network with an average erasure probability of $\epsilon =0.1$, by means of using the methods presented earlier in this tutorial, namely, the optimal method of \cite{16513519}, the efficient heuristic in \algref{alg2}, and the numerical binary swarm optimization method. \fref{fig:rwy} particularly highlights how both the optimal clique search algorithm and the particle swarm optimization algorithm outperform the heuristic algorithm (i.e., Alg.~1). The relative satisfying performance of BPSO can be explained in this situation by the availability of a good initialization to the problem. Such behavior, however, comes at the expense of additional computational complexity as the complexity of the BPSO algorithm is $\mathcal{O}(UFTL)$ as compared with $\mathcal{O}(U^2F)$ for the simple heuristic of Alg.~1, and $\mathcal{O}(\alpha^{UF})$ for the optimal algorithm.}

As stated in \sref{sec:gra}, the complexity of solving the maximum weight clique problem depends on the number of vertices. From the construction steps above, it is clear that the graph can be as big as $UF$. Therefore, pruning the graph without compromising the maximum weight clique is of great interest. The authors in \cite{645494} suggest a pruning method that removes vertices that are surely not part of the maximum weight clique.

\subsection{Collision-Free Radio Frequency Identification Networks}

Radio Frequency IDentification (RFID) systems are wireless, and automatic identification systems composed of tags and readers. Thanks to their low production costs, fast deployment, reusability, and accuracy, RFID systems have been employed in different sectors including security systems, supply chain management, retail marketing, smart university, and identification of products at check-out points, e.g., see \cite{finkenzeller2010rfid,floerkemeier2007rfid,ni2003landmarc} and references therein.

A typical RFID network is composed of one or multiple wireless readers. The identification (ID) is carried by a ``tag" that contains an integrated circuit chip with an antenna. The RFID readers can read the tag's identification and other related information within the interrogation range. This is achieved by the tag's antenna that captures the electromagnetic energy from the reader and feeds it into the integrated circuit to output the appropriate information. Furthermore, due to constraints on the processing time and link-layer protocols, the number of tags read by a reader is often bounded. Therefore, the RFID reader is limited by a maximum number of tags that can be read and a maximum interrogation range \cite{sandoval2005mobile,tran2009probabilistic}.

Multiple studies have been dedicated to deploy and activate readers such that all tags within the workspace can be read. However, in practical scenarios, some tags may not be accessible due to reader-to-reader or/and reader-to-tag collisions as multiple RFID readers can be deployed in the same geographic area. Such interference can be mitigated by optimizing the interrogation ranges of the RFID readers. The problem of activating the RFID readers and adjusting their interrogation ranges to cover maximum tags without collisions is known as the Reader-Coverage Collision Avoidance Arrangement (RCCAA) problem. The rest of the section illustrates the use of the weighted-clique problem in solving the RCCAA problem.

\subsubsection{System Model and Parameters}

An RFID system is composed of a set $\mathcal{T} = \{t_1,\ t_2,\ \cdots ,\ t_n\}$ of RFID transponders, known as tags, and $\mathcal{R} = \{r_1,\ r_2,\ \cdots ,\ r_m\}$ of RFID transceivers, known as readers. As stated earlier, readers are responsible for transmitting signals to and receiving responses from the tags within their interrogation ranges. This tutorial assumes that each reader has a discrete number $\epsilon$ of interrogation ranges, denoted by $\mathcal{D} = \{ d_1,\ d_2,\ \cdots ,\ d_\epsilon\}$, but can only use one of these $\epsilon$ interrogation ranges at each time slot. Furthermore, each reader can access a maximum of $\alpha$ tags due to energy and processing time constraints.

\textcolor{black}{To avoid reader-to-tag collisions in the above-described RFID system, each tag should be within the interrogation range of at most a single reader. Similarly, to avoid reader-to-reader collisions, an activated reader cannot be within the interrogation range of another activated reader. Therefore, given the above system description, the RCCAA problem consists of adjusting the readers' interrogation ranges so as to cover the maximum number of tags without collisions.}

The above-mentioned RCCAA problem is NP-hard \cite{7321778}. Indeed, one can show that the reduced RCCAA problem wherein $\epsilon =1$, i.e., fixed interrogation radius, requires solving the exact cover by 3-sets (X3C) problem, which is NP-hard \cite{garey2002computers}. The next part shows that the RCCAA problem is equivalent to a maximum weight independent set search, which can be solved more efficiently than generic integer programs.

\subsubsection{Graph Construction and Problem Reformulation}

Let $\gamma_r^d \subseteq \mathcal{T}$ represent the set of tags that can be read by the $r$-th reader assigned the interrogation radius $r$, for all readers $r \in \mathcal{R}$ and radius $d \in \mathcal{D}$. Consider the undirected graph $\mathcal{G}(\mathcal{V},\mathcal{E})$ wherein a vertex $v_r^d$ is generated for each feasible combination of reader and interrogation radius, i.e., $|\gamma_r^d| \leq \alpha$.

From the system constraints described above, collisions are avoided by insisting that each tag should be within the interrogation range of at most a single reader and that the intersection of interrogation regions of active readers is empty. These conditions translate to connectivity constraints in the above graph as follows. An edge is generated between nodes $v_r^d$ and $v_{r^\prime}^{d^\prime}$ if and only if $\gamma_r^d \cap \gamma_{r^\prime}^{d^\prime} = \varnothing$.

Given the graph construction as above, it can be noted that all independent sets in the graph correspond to a feasible solution to the RCCAA problem. Finally, by setting the weight of each vertex $v_r^d$ to $|\gamma_r^d|$, for all readers $r \in \mathcal{R}$ and radius $d \in \mathcal{D}$, the weight of the independent set corresponds to the number of tags that can be read. In other words, the RCCAA problem is equivalent to a maximum weight independent set search in the graph above.

The graphical solution presented above is first derived in \cite{7321778}, which solves the RCCAA problem of activating readers and adjusting their interrogation ranges to cover maximum tags without collisions, subject to the limited number of tags read by a reader. The framework is extended in a number of studies, e.g., \cite{jhuang2017new,nguyen2017mobile,wang2017automatic,tuba2017multi,8327107}, to fit multiple communication and usage paradigms. For example, the authors in \cite{8327107} consider the RCCAA problem in which both the interrogation and interference ranges are optimized. The problem is solved using a maximum weight independent set in an appropriately constructed graph that generalizes the RCCAA graph of \cite{7321778}.

\section{Applications of the $k$-clique Problem in Communications and Signal Processing} \label{sec:gra4}

\textcolor{black}{As seen in the previous sections, the maximum and maximum weight clique problems have numerous applications in communications and signal processing. The applications mentioned in the past two sections do not impose any constraints on the size of the clique under study. The paper now provides size-constrained clique applications and shows how they can be solved using the maximum $k$-clique and the maximum weight $k$-clique techniques. These applications include social network analysis \cite{balasundaram2011clique}, community detection \cite{krebs2002mapping,lancichinetti2009community,8425876,8088014}, network monitoring \cite{7910269}, and bioinformatics and biomedicine \cite{6732487}.}

\textcolor{black}{The maximum $k$-clique algorithm is a classical algorithm that is well adapted for clustering and community detection by detecting cohesive subgraphs. However, as noted in \sref{sec:gra}, the approach fails in the presence of noise due to the strict requirement of cliques. Besides the concept of para-clique introduced in \sref{sec:gra}, Seidman and Foster \cite{seidman1978graph} introduce $k$-plexes as a degree-based relaxation of graph completeness. Inspired by $k$-clique algorithms, the authors in \cite{balasundaram2011clique} design an exact BnB algorithm for the $k$-plex problem and demonstrate its effectiveness in globally solving the problem on massive and sparse graphs. More information on the $k$-plex problem and the available algorithms can be found in \cite{mcclosky2012combinatorial}.}

The $k$-clique formulation is particularly interesting for identifying communities within a network with various applications across multiple fields, e.g., homeland security, network monitoring, and disease detection. For example, while the authors in \cite{krebs2002mapping} suggest using the maximum $k$-clique algorithm to identify terrorist cells, reference \cite{6732487} exploits the algorithm to identify type II diabetes from the gene expression. A comparative study of the different approaches and algorithms for community detection is available in \cite{lancichinetti2009community} and \cite{8425876}.

The rest of the section investigates in full detail the maximum $k$-clique techniques to solve the resource allocation problem in cloud radio access networks (CRANs) with connectivity constraints. While the section focuses on cloud radio access networks, the proposed resource allocation technique can fit various other problems in communications and signal processing of similar structure.

\subsection{Resource Allocations in Cloud-Radio Access Networks}

The ever-increasing number of connected devices combined with their tremendous demand for data-hungry services strain today's wireless networks. The problem is expected to further escalate for the next generation wireless systems (5G and beyond) \cite{6824752}, due to the planned massive deployment of small cells. Furthermore, the progressive move towards full-spectrum reuse makes large-scale interference management among the different transmitters a necessity. Such interference management of all base-stations (BSs) from different tiers and of various sizes requires adopting a CRAN architecture. By connecting all BSs to a central computing unit, i.e., the cloud, CRANs offer a practical platform for the implementation of coordinated systems and thus an efficient and reliable interference mitigation mechanism.

This part addresses the interference mitigation problem in CRANs \cite{7143328} and proposes the optimal scheduling policy through a $k$-clique-based solution. In this context, scheduling denotes the strategy according to which the cloud assigns users to the different time/frequency radio resource blocks (RRBs) at each BS (also named as remote radio head (RRH)). In the classical wireless networks literature, such scheduling is often performed without requiring inter-BS coordination by assuming a pre-known association of users and base-stations, e.g., the classical proportionally fair scheduling \cite{6525475,5062043} pre-assigns users to BSs in a way that guarantees fairness. In contrast, by synchronizing the transmit frames of the connected RRHs, the cloud is capable of performing network-wide scheduling, which results in more efficient use of the radio resources, and thus in an improvement in the network performance.

This subsection solves the scheduling problem in CRANs by introducing a graph representation of all possible associations between users, remote radio heads, and their available radio resources, and then by casting the problem as a maximum weight $k$-clique. The next two parts introduce the considered system model and formulate the scheduling problem. Afterward, the scheduling graph is designed in such a way that each clique of a given size represents a feasible schedule. The problem is then formulated as a search for the maximum weight clique of a given size, i.e., a $k$-clique problem.

\subsubsection{Scheduling-Level Coordination}

\begin{figure}[t]
\centering
\includegraphics[width=0.6\linewidth]{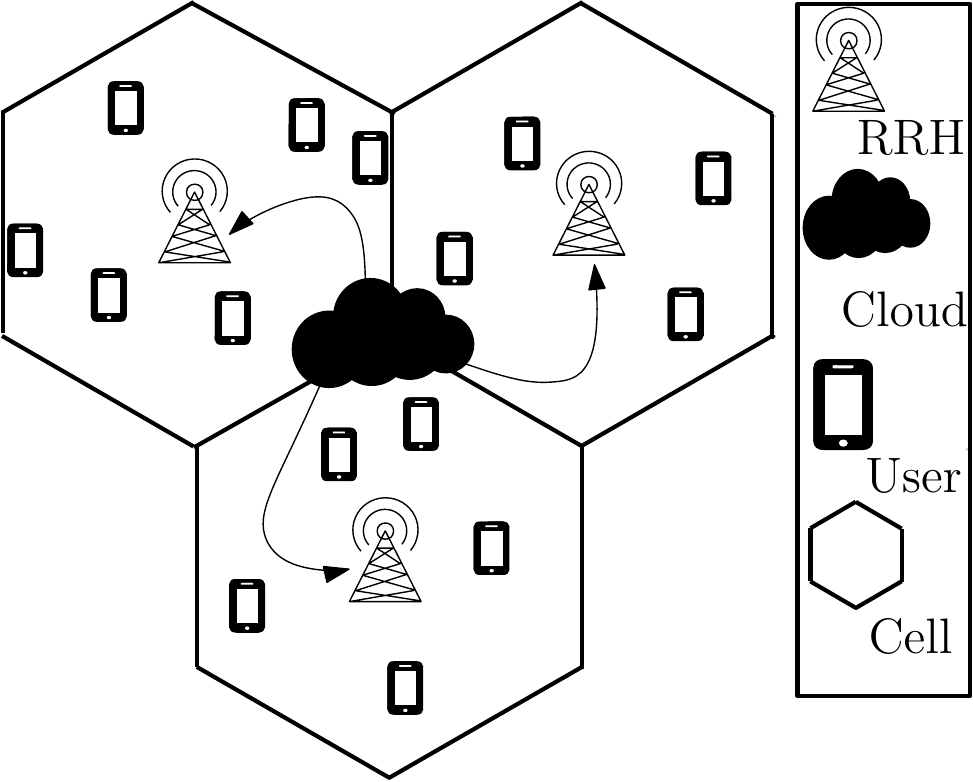}\\
\caption{A cloud radio access network comprising $3$ base-stations seen as remote radio heads and serving a total of $15$ users. The cloud communicate with the RRHs through low-rate backhaul/fronthaul links.}\label{fig:network}
\end{figure}

\textcolor{black}{As stated previously, the cloud coordinates the different transmit frames of the connected remote radio heads. The coordination level of these radio resources depends primarily on the capacity of the backhaul/fronthaul links, i.e., the capacity of the links connecting the RRHs to the cloud. As a consequence, different coordination strategies are possible in C-RANs, e.g., the signal-level, the scheduling-level, and a hybrid coordination \cite{3514852}.}

The signal-level coordination \cite{6831362,6786060,6588350,117665} performs joint encoding and decoding of the users' data through the various RRHs in the network. As such, signal-level coordination requires sharing the users' data streams among all (or a subset of) the connected RRHs, which necessitates high-capacity, low-latency backhaul links. Furthermore, sophisticated compression algorithms are required to address the quantization noise inherent to the finite capacity of the links.

\textcolor{black}{Under scheduling-level coordination, on the other hand, the cloud mitigates interference by assigning users to the radio resource blocks of the different RRHs, under the system limitation that each user can be connected to at most a single RRH. Such a constraint voids the need to share the users' data streams among the multiple RRHs, which is more suitable with networks with low data-rate backhaul links, e.g., wireless backhaul links. This paper focuses on scheduling-level coordination schemes \cite{6811617,117665}, as they offer simplified, yet efficient, resource coordination frameworks.}

\subsubsection{Coordinated Scheduling in CRANs}

\begin{figure}[t]
\centering
\includegraphics[width=0.6\linewidth]{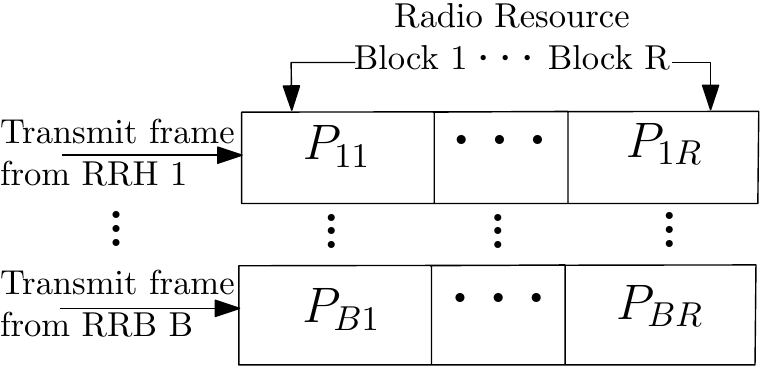}\\
\caption{Structure of the transmit frame of a cloud radio wireless network with $B$ RRHs each composed of $R$ orthogonal time/frequency RRBs. The cloud guarantees the synchronization of all transmit frames.}\label{fig:frame}
\end{figure}

Consider the downlink of a cloud wireless network in which a central computing unit, i.e., the cloud, is connected to $B$ RRHs denoted by the set $\mathcal{B}$. The network serves a set $\mathcal{U}$ of $U$ users. The RRHs and the users are equipped with single antennas. \fref{fig:network} illustrates a CRAN formed by $B$=3 BSs and $U$=15 users. The transmit frame of each RRH is composed of $R$ orthogonal time/frequency RRBs denoted by the set $\mathcal{R}$. The total number of available RRBs across all RRHs in the network is, therefore, $RB$. The transmission power of the $r$-th resource block in the $b$-th RRH is denoted by $P_{br}$ and is maintained at a fixed level. By means of control signals, the cloud guarantees the synchronization of all transmit frames, as shown in \fref{fig:frame}.

Let $h_{br}^{u} \in \mathds{C}$ be the channel gain from the $b$-th RRH to the $u$-th user when scheduled at the $r$-th RRB of base-station $b$'s transmit frame, for all $(u,b,r) \in \mathcal{U} \times \mathcal{B} \times \mathcal{R}$. The value of these channel gains are assumed both to be known at the cloud through dedicated training sequences, and to remain constant during the transmission period of one transmit frame. The signal-to-interference plus noise-ratio experienced by the $u$-th user when served by the $r$-th RRB of the $b$-th RRH frame can then be expressed as follows:
\begin{align}
\text{SINR}_{br}^{u} = \cfrac{P_{br} |h_{br}^{u}|^2}{\Gamma\left(\sigma^2+ \sum\limits_{b^\prime \neq b}P_{b^\prime r}|h_{b^\prime r}^{u}|^2\right)}, \label{eqPo}
\end{align}
where $\sigma^2$ is the Gaussian thermal noise variance, and $\Gamma$ denotes the SINR gap from Shannon capacity which accounts for the use of finite length codewords and practical constellations. Let $\text{R}_{br}^{u}$ be the data-rate of the $u$-th user when served by the $r$-th RRB of the $b$-th RRH frame, written as:
\begin{align}
\text{R}_{br}^{u} = \log_2 \big( 1+\text{SINR}^{u}_{br}\big). \label{eqrate}
\end{align}
Note that the SINR expression in \eref{eqPo} shows that the interference at the $r$-th RRB in the $b$-th RRH is seen only from RRBs with the same index in the other RRH, which is closely coupled with the synchronization capabilities of the cloud.

\textcolor{black}{Under the scheduling-level mode, the cloud aims to maximize the system throughput by scheduling users to RRBs under the following constraints:
\begin{itemize}
\item \textbf{C}1: Each user can connect at most to one RRH, but possibly to multiple RRBs of that RRH transmit frame.
\item \textbf{C}2: Each RRB should serve one and exactly one user.
\end{itemize}}

Define $X_{ubr}$ as the binary variable which denotes whether the $u$-th user is served by the $r$-th RRB of the $b$-th RRH or not. Similarly, define $Y_{ub}$ as the binary variable which denotes whether the $u$-th user is served by the the $b$-th RRH or not. The weighted sum-rate maximization problem of interest can then be formulated as follows:
\begin{subequations}
\label{Original_optimization_problem}
\begin{align}
\max_{X_{ubr},Y_{ub}} & \sum_{u,b,r} X_{ubr}\pi_{ubr}\log_2 \big( 1+\text{SINR}^{u}_{br}\big) \\
\label{constraint2}  {\rm s.t.\ }  &Y_{ub} = \min\bigg(\sum_{r}X_{ubr},1\bigg), \forall \ (u,b)\in\mathcal{U} \times \mathcal{B},\\
\label{constraint1} & \sum_{b}Y_{ub}\leq 1,\quad \forall\ u\in\mathcal{U},\\
\label{constraint3} & \sum_{u}X_{ubr}=1, \quad\forall\ (b,r)\in\mathcal{B}\times\mathcal{R},
\end{align}
\end{subequations}
where the optimization is over the binary variables $X_{ubz}$, $Y_{ub}$, and where $\pi_{ubr}$ are fixed weights which are used to scale the rate terms, so as to either provide user fairness, or to prohibit users from connecting to specific RRBs and/or RRHs. In the optimization problem \eref{Original_optimization_problem}, constraints \eref{constraint2} and \eref{constraint1} translate the system constraint \textbf{C}1. Similarly, the constraint \eref{constraint3} corresponds to system constraint \textbf{C}2.

\subsubsection{User Scheduling via $k$-clique Search}

In order to solve problem \eref{Original_optimization_problem}, this section utilizes a graph-theory based solution by first building an illustrative graph, known as the scheduling graph, where all feasible schedules are represented by cliques of size $RB$. To that end, first define the set of associations $\mathcal{A} = \mathcal{U} \times \mathcal{B} \times \mathcal{R}$ such that each association $a \in \mathcal{A}$ represents an association of a particular user to a particular RRB of some RRH frame. Reciprocally, let the mapping $\varphi_u$, $\varphi_b$, and $\varphi_r$ from the set of associations $\mathcal{A}$ to $\mathcal{U}$, $\mathcal{B}$, and $\mathcal{R}$, respectively. In other terms, for $a =(u,b,r) \in \mathcal{A}$, we have $\varphi_u(a)=u$, $\varphi_b(a)=b$, and $\varphi_r(a)=r$.

The scheduling graph $\mathcal{G}(\mathcal{V},\mathcal{E})$ is then constructed by generating a vertex $v \in \mathcal{V}$ for each association $a \in \mathcal{A}$. Edges are generated in such a way that connected vertices represent a partially feasible schedule, i.e., the connected vertices are not conflicting. Using the canonical identification $\mathcal{V} = \mathcal{A}$, an edge $\epsilon \in \mathcal{E}$ is generated between the vertices $v$ and $v^\prime$ if both the following connectivity conditions hold:
\begin{itemize}
\item \textbf{CC}1: if $\varphi_u(v) = \varphi_u(v^\prime)$ then $\varphi_b(v) = \varphi_b(v^\prime)$. This condition translates the fact that each user cannot connect to multiple RRHs.
\item \textbf{CC}2: $(\varphi_b(v),\varphi_z(v)) \neq (\varphi_b(v^\prime),\varphi_z(v^\prime))$. This condition affirms the same RRB can be assigned to at most a single user.
\end{itemize}

\begin{figure}[t]
\centering
  \includegraphics[width=.6\linewidth]{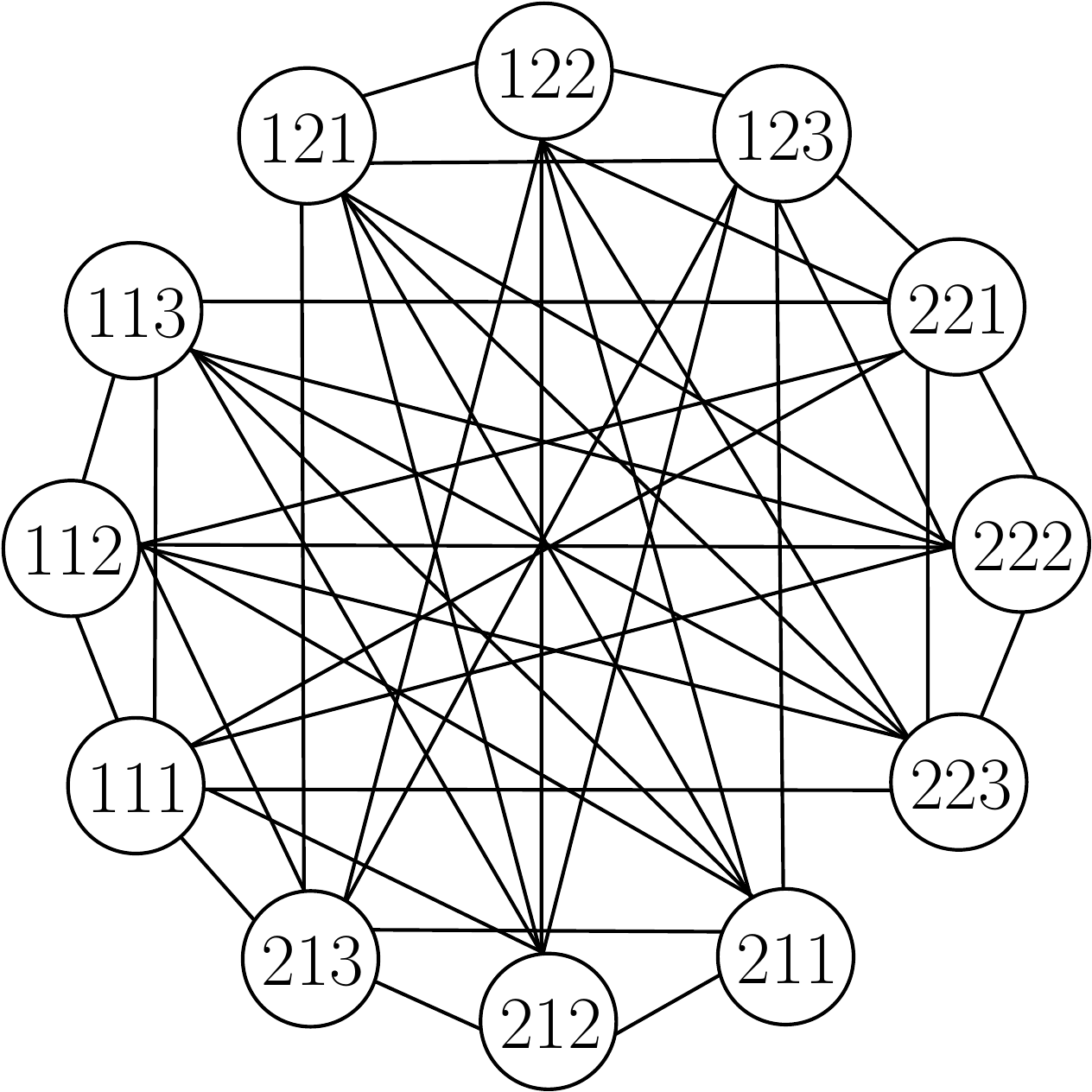}\\
  \caption{Example of scheduling graph for cloud radio access network comprising $U=2$ users, $B=2$ RRHs each containing $R=3$ RRBs. Each vertex in the graph is labelled $UBR$ wherein $U$, $B$, and $R$ refer to the index of the user, RRH, and RRB, respectively.}\label{fig:graph}
\end{figure}

As stated earlier, the graph is constructed such as each clique of size $RB$ represents a feasible schedule (see \cite{3514852}). Therefore, by assigning to each vertex $v \in \mathcal{A}$ the weight $w(v) = \pi_{ubr}X_{ubr}\log_2 \big( 1+\text{SINR}^{u}_{br}\big)$ for $(\varphi_u(v),\varphi_b(v),\varphi_r(v)) = (u,b,r)$, it becomes clear that the weight of each clique coincides with the objective function in the optimization problem \eref{Original_optimization_problem}. Finally, one can conclude that the optimal schedule corresponds the the maximum weight $RB$-clique in the scheduling graph which can be efficiently solved using state of the art graph theory techniques. An example of the scheduling graph is provided in \fref{fig:graph} for a network containing $U=2$ users, $B=2$ RRHs, and $R=3$ RRBs. {There are only two possible cliques of size $B \times R=2 \times 3 =6$ which are $\{111,112,113,221,222,223\}$ and $\{121,122,123,211,212,213\}$.}

\textcolor{black}{Note that by construction, the graph connectivity conditions \textbf{CC}1 and \textbf{CC}2 are a direct consequence of the system constraints \textbf{C}1 and \textbf{C}2. As a matter of fact, the authors in \cite{8027131} discuss how different system connectivity constraints, and thus different coordination levels, give various types of problem reformulations, and consequently different graphical formulations. Therefore, depending on the wireless backhaul condition, the cloud can take advantage of the different types of approaches to dynamically adapt the coordination level for better overall system performance. However, such gain comes at the expense of a relative increase in the algorithmic computational complexity, which can be reduced under an additional assumption on the channel model \cite{douik2017low}.}

\section{\textcolor{black}{Clique Problems and Mixed Discrete-Continuous Optimization Problems}} \label{sec:gra5}

\textcolor{black}{The above parts of the manuscript present a unique blend of clique-based techniques which serve to solve several contemporary discrete optimization problems in communications and signal processing. To provide additional perspectives to our tutorial, the paper now provides insights about how is the clique problem useful in dealing with mixed discrete-continuous problems. In particular, the manuscript illustrates the use of the clique formulation to solve a mixed continuous-integer program while addressing the resource allocation problem in cloud radio access networks.}

\textcolor{black}{All solutions presented thus far solely focus on solving discrete optimization problems using clique problem reformulations. Such techniques can be further utilized in mixed continuous-discrete optimization problems, e.g., see \cite{zhai2018spectrum,8365765}. More specifically, the tutorial now focuses on the joint coordinated scheduling and power control problem in CRANs \cite{7342981}.} Under this scenario, the cloud is responsible for both scheduling users at the RRBs in the RRHs' frames, and also determining the power levels of each RRB. Let $P_{br}$ and $P^{\max}_{br}$ be the power level and the maximum available power at the $r$-th RRB of the $b$-th RRH frame, respectively. Furthermore, let $\mathbf{P} = [P_{br}]$ be the matrix that contains all the power levels. The joint coordinated scheduling and power control problem subject to the system constraints \textbf{C}1 and \textbf{C}2 can be written as follows:
\begin{subequations}
\label{Original_optimization_problem2}
\begin{align}
\max_{X_{ubr},Y_{ub}} & \sum_{u,b,r}\pi_{ubr}X_{ubr}\log_2 \big( 1+\text{SINR}^{u}_{br}(\mathbf{P})\big) \\
 {\rm s.t.\ }  &Y_{ub} = \min\bigg(\sum_{r}X_{ubr},1\bigg), \forall \ (u,b)\in\mathcal{U} \times \mathcal{B},\\
& \sum_{b}Y_{ub}\leq 1,\quad \forall\ u\in\mathcal{U},\\
& \sum_{u}X_{ubr}=1, \quad\forall\ (b,r)\in\mathcal{B}\times\mathcal{R},\\
\label{constraint6} & P_{br} \leq P^{\max}_{br} , \forall\ (b,r)\in\mathcal{B}\times\mathcal{R},
\end{align}
\end{subequations}
where the optimization is over the binary variables $X_{ubz}$, $Y_{ub}$, and over the continuous variables $P_{br}$, $\ \forall \ (b,r) \in \mathcal{B} \times \mathcal{R}$. The problem formulation \eref{Original_optimization_problem2} is similar to the one presented in \eref{Original_optimization_problem}, except that the power levels of all the RRBs become now among the optimization variables. The limitation on the available power at each RRB is also appended through the system constraint \eref{constraint6}.

\textcolor{black}{In order to solve the joint scheduling and power allocation problem \eref{Original_optimization_problem2}, reference \cite{7342981} suggests combining continuous optimization to derive the weights of the vertices and clique optimization to solve the resulting clique problem. This is done by constructing a local power control graph for each RRB index $r$, in which each vertex represents $B$ associations for the $r$-th RRB and its weights reflect their contribution to the network obtained from the power levels. The power levels of these $B$ associations can be determined by using either optimal, e.g., \cite{4801507}, or efficient, e.g., \cite{52165132}, power allocation algorithms resulting in optimal and heuristic joint allocation, respectively.}

\begin{figure}[t]
\centering
  \includegraphics[width=1\linewidth]{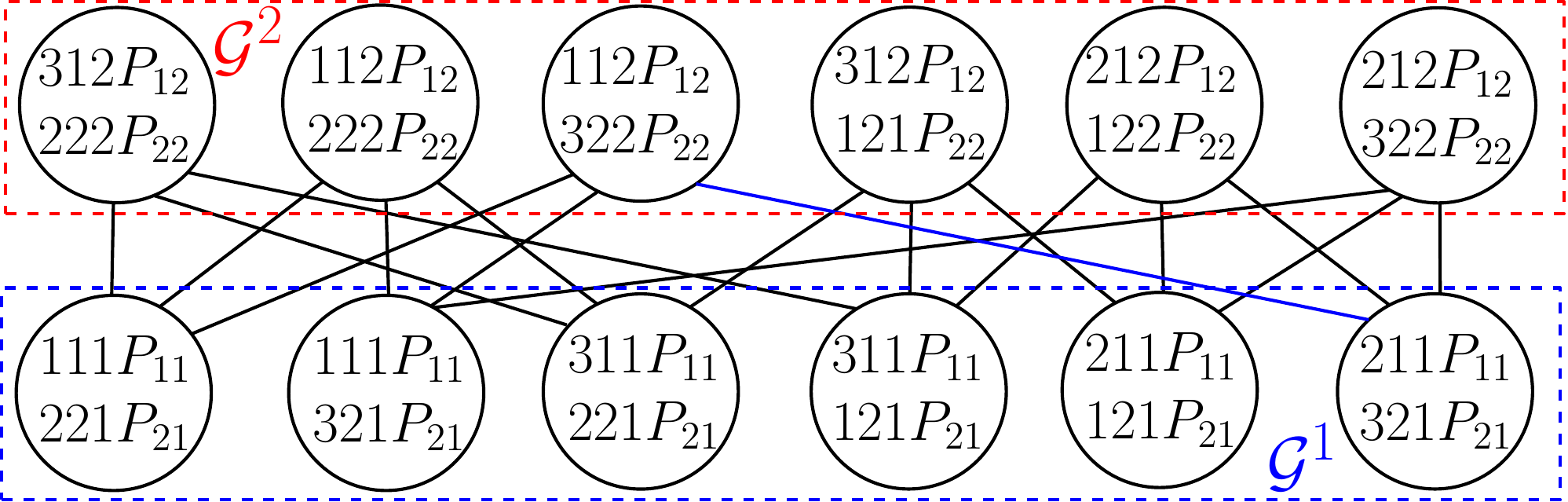}\\
  \caption{Joint scheduling and power control graph for a network composed of of $2$ base-stations, $2$ power-zones per base-station and $3$ users. The local power control graph are represented by $\mathcal{G}_1$ and $\mathcal{G}_2$.}\label{fig:net}
\end{figure}

The joint scheduling and power allocation graph $\mathcal{G}(\mathcal{V},\mathcal{E})$ is afterwards generated by taking the union of all local power control graphs. Two vertices $v$ and $v^\prime$ belonging to \emph{distinct} local power control graphs are connected by an edge if their combination results in a feasible schedule. In other words, vertices $v$ and $v^\prime$ are connected if the association they represent does not include users connected to multiple RRHs. This can be mathematically represented $\forall \ a \in v, \ \forall \ a^\prime \in v^\prime$ by the following condition:
\begin{align}\label{eq14}
&\delta( \varphi_u(a) - \varphi_u(a^\prime))\delta( \varphi_b(a) - \varphi_b(a^\prime))= \delta( \varphi_u(a) - \varphi_u(a^\prime)), \nonumber
\end{align}
wherein $\delta(.)$ is the discrete Dirac function that is equal to $1$ if its argument is $0$, and $0$ otherwise. \fref{fig:net} illustrates the joint scheduling and power control graph for a CRAN composed of $U=3$ users, $B=2$ RRHs each containing $R=2$ RRBs. The local power control graphs of RRB $1$ and $2$ are represented by $\mathcal{G}_1$ and $\mathcal{G}_2$, respectively.

After constructing the joint scheduling and power allocation graph as above, the optimal solution to the mixed discrete and continuous optimization problem \eref{Original_optimization_problem2} is given by the maximum weight $R$-clique wherein the weight of each vertex $v$ is defined as:
\begin{align}
w(v) = \sum_{a=(u,b,r) \in v} \pi_{ubr}X_{ubr}\log_2 \big( 1+\text{SINR}^{u}_{br}(\mathbf{P})\big)
\end{align}

\vspace{-0.3cm} 

\section{Conclusion} \label{sec:con}

\textcolor{black}{Graph theory provides a robust set of tools for unveiling the structure of multiple communications and signal processing problems. This tutorial focuses on a particular graph theory problem, known as the clique problem, and illustrates its use through numerous contemporary examples in communications and signal processing. Such techniques are expected to further play a crucial role in future applications given the continuous shift toward fully digital systems yielding many more interesting insights and results, especially for clique-based mixed discrete-continuous optimization. To the best of the authors' knowledge, this is the first tutorial which presents the clique search problem as a crucial technique for optimizing communications and signal processing systems. The tutorial does not only shed light on timely discrete optimization problems in the field but also presents a promising framework for solving futuristic problems of similar structures.}

\vspace{-0.1cm} 

\section*{Acknowledgment}

\textcolor{black}{The authors wish to thank Ahmed K. Sultan Salem for his helpful discussions and valuable comments. They further extend their gratitude to the anonymous reviewers whose comments and suggestions helped improve the quality of the manuscript.}

\vspace{-0.1cm} 

\bibliographystyle{IEEEtran}
\bibliography{citations}

\vspace{-02cm} 

\begin{IEEEbiography}[{\includegraphics[width=1in,height=1.25in,clip,keepaspectratio]{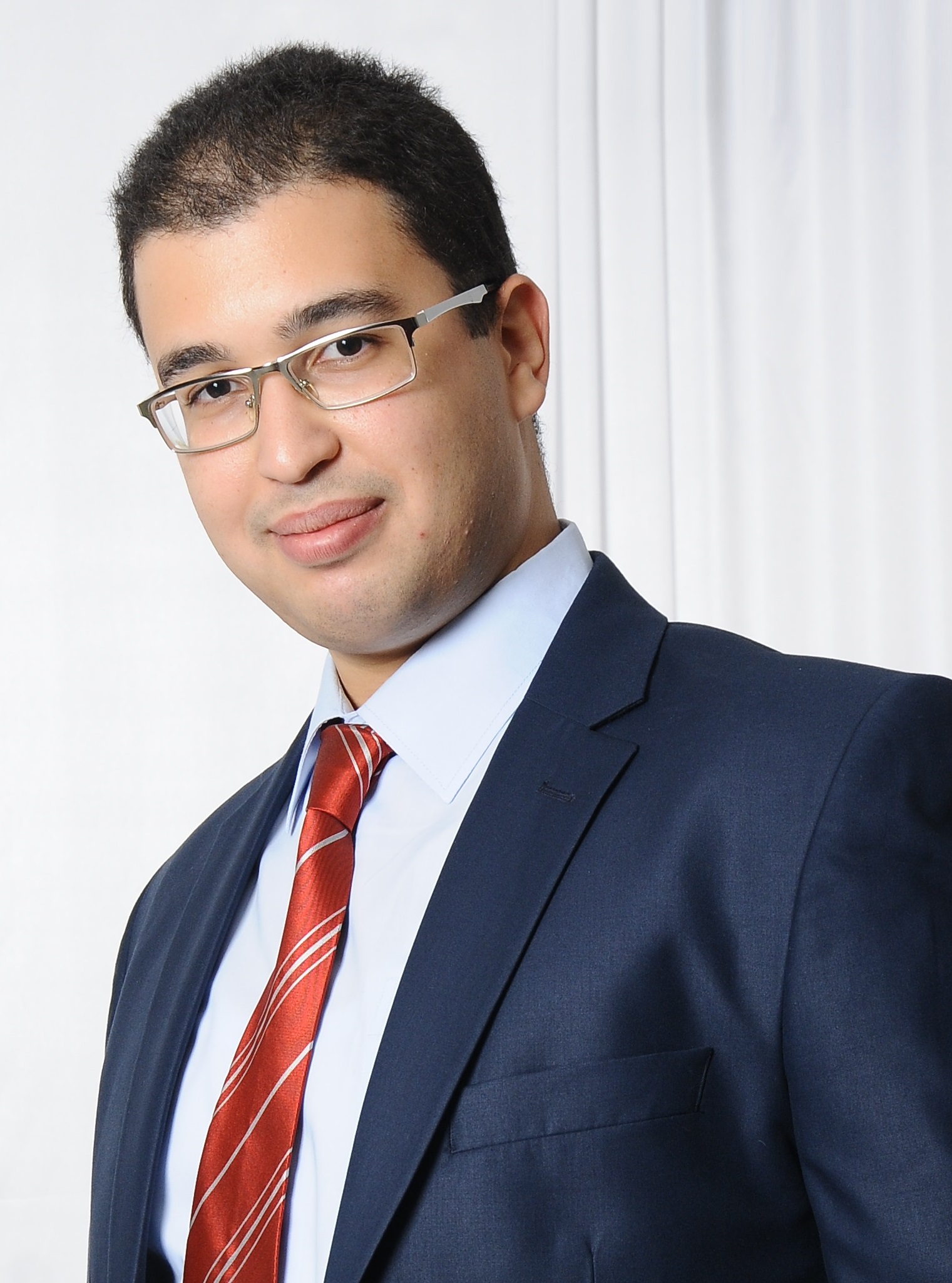}}]{Ahmed Douik}(S'13)
received the Eng. degree in electronic and communication engineering (with first class honors) from the Ecole Polytechnique de Tunisie, Tunisia, in 2013, the M.S. degree in electrical engineering from King Abdullah University of Science and Technology, Thuwal, Saudi Arabia, in 2015. He is now pursuing his Ph.D. at the California Institute of Technology, Pasadena, CA, USA. His research interests include cloud-radio access networks, network coding, single and multi-hop transmissions, and cooperation communication.
\end{IEEEbiography}

\begin{IEEEbiography}[{\includegraphics[width=1in,height=1.25in,clip,keepaspectratio]{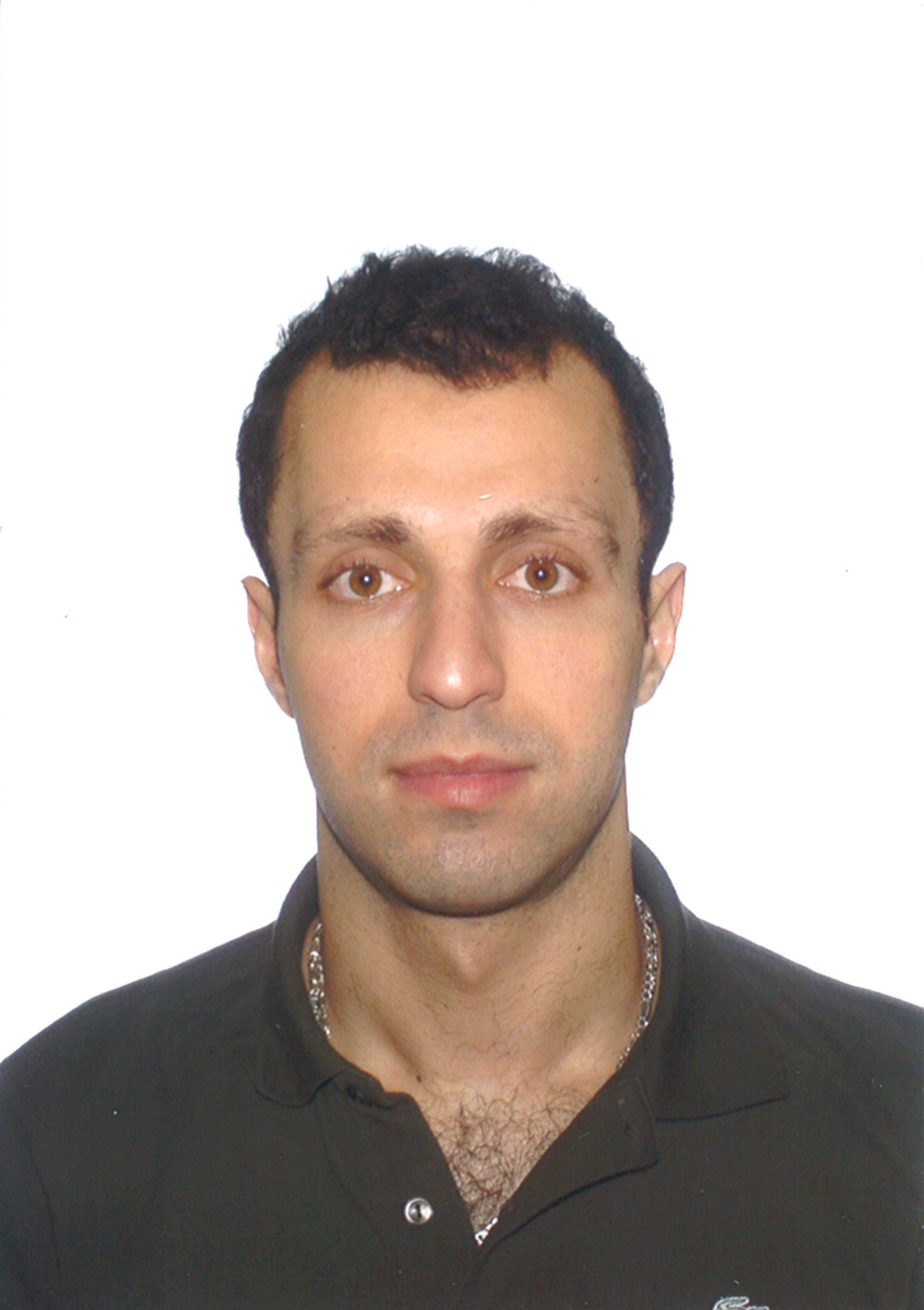}}]{Hayssam Dahrouj}(S'02, M'11, SM'15)
received his B.E. degree (with high distinction) in computer and communications engineering from the American University of Beirut (AUB), Lebanon, in 2005, and his Ph.D. degree in electrical and computer engineering from the University of Toronto (UofT), Canada, in 2010. In May 2015, he joined the Department of Electrical and Computer Engineering at Effat University as an assistant professor, and also became a visiting scholar at King Abdullah University of Science and Technology (KAUST). Between April 2014 and May 2015, he was with the Computer, Electrical and Mathematical Sciences and Engineering group at KAUST as a research associate. Prior to joining KAUST, he was an industrial postdoctoral fellow at UofT, in collaboration with BLiNQ Networks Inc., Kanata, Canada, where he worked on developing practical solutions for the design of non-line-of sight wireless backhaul networks. His contributions to the field led to five patents. During his doctoral studies at UofT, he pioneered the idea of coordinated beamforming as a means of minimizing intercell interference across multiple base stations. The journal paper on this subject was ranked second in the 2013 IEEE Marconi paper awards in wireless communications. His main research interests include cloud radio access networks, cross-layer optimization, cooperative networks, convex optimization, distributed algorithms, and free-space optical communications.
\end{IEEEbiography}

\begin{IEEEbiography}[{\includegraphics[width=1in,height=1.25in,clip,keepaspectratio]{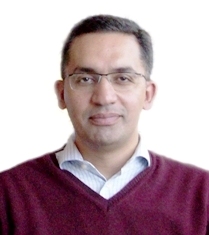}}]{Tareq Y. Al-Naffouri}(M'10)
Tareq Al-Naffouri received the B.S. degrees in mathematics and electrical engineering (with first honors) from King Fahd University of Petroleum and Minerals, Dhahran, Saudi Arabia, the M.S. degree in electrical engineering from the Georgia Institute of Technology, Atlanta, in 1998, and the Ph.D. degree in electrical engineering from Stanford University, Stanford, CA, in 2004.

He was a visiting scholar at California Institute of Technology, Pasadena, CA, from January to August 2005 and during summer 2006. He was a Fulbright scholar at the University of Southern California from February to September 2008. He has held internship positions at NEC Research Labs, Tokyo, Japan, in 1998, Adaptive Systems Lab, University of California at Los Angeles in 1999, National Semiconductor, Santa Clara, CA, in 2001 and 2002, and Beceem Communications Santa Clara, CA, in 2004. He is currently an Associate at the Electrical Engineering Department, King Abdullah University of Science and Technology (KAUST). His research interests lie in the areas of sparse, adaptive, and statistical signal processing and their applications and in network information theory.  He has over 150 publications in journal and conference proceedings, 9 standard contributions, 10 issued patents, and 6 pending.

Dr. Al-Naffouri is the recipient of  the IEEE Education Society Chapter Achievement Award in 2008 and Al-Marai Award for innovative research in communication in 2009. Dr. Al-Naffouri has also been serving as an Associate Editor of Transactions on Signal Processing since August 2013.
\end{IEEEbiography}

\begin{IEEEbiography}[{\includegraphics[width=1in,height=1.25in,clip,keepaspectratio]{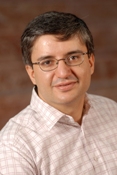}}]{Mohamed-Slim Alouini}(S'94-M'98-SM'03-F'09)
was born in Tunis, Tunisia. He received the Ph.D. degree in electrical engineering from the California Institute of Technology (Caltech), Pasadena, CA, USA, in 1998. He served as a faculty member at the University of Minnesota, Minneapolis, MN, USA, then in the Texas A\&M University at Qatar, Education City, Doha, Qatar, before joining King Abdullah University of Science and Technology (KAUST), Thuwal, Makkah Province, Saudi Arabia, as a Professor of electrical engineering in 2009. His current research interests include the modeling, design, and performance analysis of wireless communication systems.
\end{IEEEbiography}

\end{document}